\documentclass[12pt]{article}

\usepackage{amsmath,amssymb}
\usepackage[dvips]{graphicx}

\title{Einstein and the Early Theory of Superconductivity, 1919--1922}

\author{Tilman Sauer\\
	{\footnotesize Einstein Papers Project}\\[-0.20cm]
	{\footnotesize California Institute of Technology 20-7}\\[-0.20cm]
	{\footnotesize Pasadena, CA~91125, USA}\\[-0.20cm]
	{\footnotesize tilman@einstein.caltech.edu}}

\date{}
\vspace{-1.5cm}
\begin{document}

\maketitle

\renewcommand{\baselinestretch}{0.95}
\begin{abstract}
Einstein's early thoughts about superconductivity are discussed
as a case study of how theoretical physics reacts to experimental findings
that are incompatible with established theoretical notions. One such notion that
is discussed is the model of electric conductivity implied by Drude's electron
theory of metals, and the derivation of the Wiedemann-Franz law within this framework.
After summarizing the experimental knowledge on superconductivity around 1920,
the topic is then discussed both on a phenomenological level in terms of
implications of Maxwell's equations for the case of infinite conductivity,
and on a microscopic level in terms of suggested models for superconductive charge transport. 
Analyzing Einstein's manuscripts and correspondence as well as his own
1922 paper on the subject, it is shown that Einstein had a sustained interest
in superconductivity and was well informed about the phenomenon. 
It is argued that his appointment as special professor
in Leiden in 1920 was motivated to a considerable extent by his perception as a leading
theoretician of quantum theory and condensed matter physics and the hope that he
would contribute to the theoretical direction of the experiments done at Kamerlingh
Onnes' cryogenic laboratory. Einstein tried to live up to these expectations
by proposing at least three experiments on the phenomenon, one of which was
carried out twice in Leiden. Compared to other theoretical proposals at the time, 
the prominent role of quantum concepts was characteristic of Einstein's understanding 
of the phenomenon. The paper concludes with comments on
Einstein's epistemological reflections on the problem.
\end{abstract}

\clearpage

\renewcommand{\baselinestretch}{1.0}
\vspace{-2.0cm}
{\small
\tableofcontents
}

\section*{Introduction}
\addcontentsline{toc}{section}{Introduction}

The history of superconductivity\footnote{For general historical
accounts of the experimental and theoretical developments associated
with the phenomenon of superconductivity, see \cite{Dahl1992} and
\cite{MV2003}.}  
constitutes an example of conceptual change in physics where unexpected
experimental discoveries have preceded theoretical analyses more than
once. The very discovery of superconductivity in 1911 itself is a case
in point. No definite theoretical expectations could be formulated as
to how the electric resistance would behave in the very low
temperature regime on the basis of contemporary theories and models of
electrical resistance. The sudden loss of resistivity of some metals
within a very small temperature interval at liquid helium temperatures
over several orders of magnitude to a value that was below any
experimentally observable threshold was a novelty, not expected and not
to be foreseen. The same would hold true, one may argue, for the
discovery of the Meissner effect in 1933, the second of the two
fundamental features that constitute, according to today's
understanding, the phenomenon of superconductivity. The discovery that
superconductors are perfect diamagnets and expel any magnetic fields
reversibly when entering the superconducting state, hence rendering
the superconducting state a true thermodynamic state, was made in
the context of theoretical speculations about the magnetic behavior of
supercoductors, but the effect itself was unexpected as well. 
More recently, the discovery of high-temperature superconductors by 
Bednorz and M\"uller in 1986 was again an unforeseen experimental 
discovery. Although by now many more superconductors of high
transition temperature have been identified and a wealth of detail is
known about these materials, the precise mechanism of high-temperature
superconductivity is still not yet fully understood.\footnote{For an
overview of the available experimental data and further references,
see, e.g.\ \cite{Poole2000}.}

We now know that superconductivity is a genuine macroscopic
quantum phenomenon. It defied theoretical understanding until
first successfully interpreted in terms of a macroscopic wave function
by V.L.~Ginzburg and L.D.~Landau in 1950. Similar to non-relativistic
Schr\"odinger quantum mechanics, the square of the wave function is
interpreted as a probability density for the superconducting
electrons. The wave function in the Ginzburg-Landau theory also acts
as a thermodynamic order parameter, such that the transition from
normal conductivity to the superconducting state in the absence of
magnetic fields is interpreted as a phase transition of second
order. A microscopic justification of the phenomenological
Ginzburg-Landau theory was given seven years later by J.~Bardeen,
L.N.~Cooper and J.R.~Schrieffer. In the so-called BCS theory which
constitutes today's standard explanation of superconductivity, the
macroscopic wave function is accounted for by a microscopic theory in
which an effective attractive interaction between electrons arises
from lattice phonons so that electrons associate to pairs and condense
to the macroscopic wave function. It should therefore be clear that in
the very beginning of the history of superconductivity, i.e.\ long
before the discovery of the Meissner effect and long before concepts
such as a quantum-mechanic wave function and a phase transition of second
order were available, an explanation of the phenomenon that in any sense
might come close to our modern understanding 
was well out of reach of contemporary theoreticians.

As a study of how theoretical physics is being done in practice, it is
interesting then to take a closer historical look at how physicists have
interpreted experimental data that clearly challenged the validity of
well-established concepts and theories by a phenomenon and that were, at
the same time, well out of the horizon of what could possibly be
understood at the time in any reasonable way. Gavroglu and Goudaroulis
have coined the term ``concepts out of context(s)'' to capture the
peculiar situation of theoretical attempts to come to grips with the
phenomenon of superconductivity.\footnote{\cite{GG1984},
\cite{GG1989}, see also \cite{Gavroglu1985}.} 
This paper examines one such reaction to this phenomenon, namely 
Albert Einstein's. Given Einstein's characteristic
awareness of foundational problems that allowed him not only to
overcome the limits of classical mechanics and electrodynamics with
his theories of relativity, but also to be one of the first who
perceived most clearly the limits of classical mechanics with respect
to the quantum phenomena, a reconstruction of Einstein's
interpretation of the phenomenon of superconductivity promises
insights into the theoretical horizon of the time. Indeed, as I will
show, it was Einstein who not only most clearly recognized the
challenge posed by the phenomenon to classical concepts but who also
most explicitly advocated and actively explored the use of quantum
concepts for a theoretical understanding of superconductivity.

In 1922 Einstein wrote a paper, entitled ``Theoretical remarks on the
superconductivity of metals,'' which has received comparatively little
attention from historians of
science.\footnote{\cite{Einstein1922}. B.S. Schmekel recently
published an English translation at
http://www.arxiv.org/physics/0510251 (In my quotations from this
paper, I will occasionally differ from Schmekel's translation. Except
for correspondence and writings published in the {\em Collected Papers of Albert
Einstein} where English translations were taken from the translation
volumes of this series, all other English translations are mine.). The paper
is discussed in \cite{Yavelov1980} and \cite[pp.~105--106]{Dahl1992};
see also \cite[p.~42]{MV2003}, \cite[p.~86]{Kragh1999},
\cite[p.~335]{Renn2000} for brief mentions of this paper.} One reason
for its neglect in the historical literature might be
that it appeared to be a quite isolated episode within
Einstein's published oeuvre, unconnected to his more prominently
figuring concerns. Recently, however, the editorial project of the
{\it Collected Papers of Albert Einstein} has brought to the fore some
evidence in Einstein's correspondence and unpublished manuscripts that
not only allows us to get a better picture of Einstein's thoughts on
the problem of superconductivity, but also would have us revise, or at
least nuance to some extent, our understanding of Einstein's
preoccupations.

Einstein's visit to Leiden in the fall of 1920 is mainly known for his
famous inauguration lecture on ``Ether and Relativity,'' delivered on
the occasion of his appointment as special visiting professor at
Leiden. The lecture was given on October 27, 1920, and published
separately as a little booklet
\cite{Einstein1920}.\footnote{\cite{Einstein1920}. The text of the
printed lecture was completed before 7 April 1920, and the title page
of the printed version states that the lecture was given on 5 May
1920. Due to delays in his appointment (see
\cite[pp.~xliii--xlvi]{CPAE10}), the lecture was, however, given only
on 27 October 1920, see \cite[p.~321]{CPAE7}.} Less known but equally
important at the time was the fact that Einstein spent most of his
time during this stay in Leiden in late October and early November
1920 participating in a meeting devoted to recent
developments in low temperature physics, specifically about the
problems of magnetism at low temperatures. Other participants in these
discussions included Paul Ehrenfest, Heike Kamerlingh Onnes, Willem
H. Keesom, Johannes P.\ Kuenen, Paul Langevin, Hendrik A. Lorentz, and
Pierre Weiss.\footnote{On this meeting, see \cite[p.~42]{MV2003},
\cite[p.~xlvi--xlviii]{CPAE10}. See also \cite[p.~3]{Onnes1921a}, where
the November meeting is explicitly mentioned in an introductory
footnote. In the evening edition of 25 October 1920 of the Dutch
daily {\it Nieuwe Rotterdamsche Courant} an announcement of the meeting was
published in which Einstein's role was especially emphasized: The discussions
were to center on ``the phenomena of paramagnetism at low temperatures and
the pecularities of supeconductivity. [...] The attraction of these
discussions, that will take place this week, is greatly enhanced
by the participation of Prof.~Einstein from Berlin [...]. He is in
particular participating with regards to the application of the
quanta to the explanation of the mentioned phenomena.'' The newspaper
report also mentioned that Onnes hoped to host similar meetings 
``at times that Prof.~Einstein is also in Leiden because of his 
professorship.'' I wish to thank Jeroen van Dongen for alerting me
to this newspaper article and for providing an English translation.}

In fact, the initiators of Einstein's appointment in Leiden
specifically were hoping for his input in discussions of problems in
low temperature physics. Thus Lorentz wrote to Einstein, almost a year 
earlier, on 21 December 1919:
\begin{quote}
Our Berlin colleagues will undoubtedly understand that we would like
to have you here from time to time and that, for inst., Kamerlingh
Onnes would put great store in discussing problems being addressed at
his Cryogenic Laboratory with you.\footnote{%
\cite[Doc.~229]{CPAE9}. For a very similar
comment in this spirit, see also Lorentz to Einstein, 16 January 1920
\cite[Doc.~264]{CPAE9}.}
\end{quote}
And the champion of low temperature physics himself, Kamerlingh Onnes
wrote to Einstein on 8 February 1920:
\begin{quote}
Thus, best conditions are made for stimulating investigations, guiding
ongoing analyses onto better paths, as well as exchanging fruitful
ideas of every kind. Thus, with your Leiden professorship I also
cherish the finest hopes for a flowering of the cryogenic
laboratory. Virtually no one is so closely affiliated to it as you
are. Many of the investigations performed there regard phenomena whose
relevance to quantum theory you have recognized and for whose analysis
this laboratory is somewhat of an international institution, insofar
as the area of low temperatures is concerned. So your help can bring
about much that is of benefit.

You will perhaps find me very egoistical if I already immediately ask you to
make available to me some of your precious time for devising
strategies and identifying problems. But I take that risk, dear
friend! And I assure you that I find it just as great a fortune for
the Cryogenic Laboratory as for theoretical physics that you will be
connected with Leiden as one of our own.\footnote{%
\cite[Doc.~304]{CPAE9}.}  
\end{quote} 

Although outshone by the stellar success of the 1919 confirmation of
gravitational light bending by the British eclipse expedition,
there are thus a number of indications that Einstein at the time was 
indeed considered in Leiden a leading theoretician of low temperature 
physics.\footnote{Similarly, Einstein's earlier call to Berlin in 1914 was not 
so much motivated by the expectation that he would elaborate on and expand his theory of 
special relativity
--- despite the fact that the discovery of general relativity would become
his greatest achievement of the Berlin years ---
but rather by his perception as a leading theoretician of statistical and
quantum physics and, indeed, of physical chemistry, see the discussion in 
\cite[pp.~69--77]{Renn2006}.}
What follows is an account of what we
know about Einstein's concerns with superconductivity until 1922 and 
an analysis of his interpretation of this phenomenon.

We have little evidence of Einstein's thoughts on the subject before 
1919 and, in spite of some
efforts, I did not find any comments by him from later than
1922. Nevertheless, Einstein's apparent silence on the subject may
well be attributed to our as yet insufficient knowledge of the
documents in the Einstein Archives. It is hence possible that
material of interest may come to light, e.g. in the
preparation of further volumes of Einstein's {\it Collected
Papers}. I will not address
Einstein's thoughts and ideas on other related phenomena, such as the
behavior of specific heats at low temperature, or his statistical work
that led to the identification of what we now call the Bose-Einstein
statistics. The
relative weight and significance of ideas about superconductivity was
determined to some extent by the conceptualization of normal electric
conductivity and of phenomena associated with it. But a thorough
discussion of the issue of normal metallic resisitivity, or of low
temperature phenomena in general, or of those phenomena that were at
the basis of the emergence of the new quantum theory is beyond the
scope of the present paper which, focusses exclusively on the 
problem of accounting for the phenomenon of superconductivity. 

My account will be organized as follows.  In order to address the
methodological difficulty just mentioned, my starting point will be
the canonical conceptualization of electrical and thermal
conductivity of the early twentieth century, which was based on
Drude's electron theory of metals and culminated in a quantitative
formula for the Wiedemann-Franz law.  I will discuss the standard
derivation of this law within the electron theory of metals by
annotating Einstein's own derivation as written down sketchily in his
course notes for a lecture course on the kinetic theory of heat held
in 1910. I will then give a synchronic characterization of the
state of knowledge about the phenomenon of superconductivity ca.~1920,
followed by a discussion of our evidence that Einstein was, in
fact, well informed about these experimental data through his strong
professional and personal ties to the physicists at Leiden. I will
then discuss the phenomenological theory of infinite or perfect
conductivity, as expounded in an influential contribution by Gabriel
Lippmann.  Investigations of the Maxwell equations for infinite
conductivity can be found in Paul Ehrenfest's diaries. They were also
the basis for a consideration by Einstein that was intended as
background theorizing for proposed experimental investigations of the
particular features of a Hall effect for superconductors, should such
an effect exist.  I will then turn to a discussion of microscopic
models of infinite conductivity.  In order to
provide some necessary context for Einstein's own theory, I will
discuss several contemporary proposals of microscopic charge transport
that were advanced specifically in order to meet the challenge posed
by the phenomenon of supercoductivity to the kinetic electron theory of 
metals and to account for the phenomena associated with
it. Among these are models by Johannes Stark, Frederick A.~Lindemann, Heike
Kamerlingh Onnes, Joseph John Thomson, Fritz Haber, and finally by Einstein
himself. Against the background of this horizon of theoretical
responses to the available experimental data, Einstein's own
theoretical speculations, as expounded in his only published paper on
the subject, appear as an innovative and original contribution, not the least
because he employed concepts from the emerging quantum theory.
Einstein derived testable consequences of his specific microscopic
assumptions about superconductive currents, at least one of which was
tested in Leiden by an experiment specifically designed for
this purpose. I will conclude with a discussion of
Einstein's epistemological reflections on the problem and some 
remarks on Einstein's contributions.

\section*{Drude's electron theory of metals}
\addcontentsline{toc}{section}{Drude's electron theory of metals}

At the time of the discovery of superconductivity, the electron 
theory of metals was a highly developed and sophisticated theory.\footnote{For
contemporary reviews, see \cite{Seeliger1921}, \cite{Suter1920},
\cite{Meissner1920}. For a historical discussion, see
\cite{Kaiser1987} and also \cite{HoddesonBaym1980} and \cite{Hoddesonetal1987}.
For a historical discussion of Einstein's concerns
with an electron theory of metals, see \cite{Renn2000}.} Its most
impressive success was a theoretical justification of the so-called
Wiedemann-Franz law. This law asserts that for many metals the ratio
of thermal and electrical conductivity only depends on
temperature and not on any specific properties of the metal. Part of
the success of the electron theory of metals was the fact that it
seemed to provide a well-founded and unambiguous way to also quantitatively
compute the coefficient of the temperature dependence of the Wiedemann-Franz law
also quantitatively, and that the theoretical values agreed with
reasonable accuracy with the observed values.

The model itself was extremely simple, although more detailed
theoretical discussions of its features could become quite
involved. For our purposes it will suffice to discuss its basic
features. We will do so by paraphrasing and commenting on Einstein's
own notes on a derivation of the Wiedemann-Franz law in the context of
the electron theory of metals. A brief, ``back-of-an-envelope''
derivation of this law is written down in Einstein's lecture notes
for a course on kinetic theory, held in the summer semester 1910 at
the University of Zurich.\footnote{The course notes are published as
\cite[Doc.~4]{CPAE3}. For a facsimile of the course notes, see
Einstein Archives Online (http://www.alberteinstein.info), Call
Nr. 3-003. The page dealing with the electron theory of metals is
[p.~49], i.e. \cite[pp.~232--233]{CPAE3}. For a very similar example of the
following ``back-of-an-envelope'' calculation, including the
factor-of-1/2-problem discussed below, see the first page of notes by
Niels Bohr for a lecture course on the Electron Theory of Metals, held
in 1914 at the University of Copenhagen, \cite[p.~446]{BCW1}.} In these notes, Einstein
sketched standard theoretical considerations he had obtained from his
readings of Boltzmann, Riecke, Drude, and others, as preparation for
his classes, and without explicit reference to his sources.\footnote{For
further evidence that Einstein was well acquainted with, and critical of,
contemporary research in the electron theory of metals, see Einstein
to Mileva Mari\'c, 28? May 1901. In this letter, he reports about having 
read \cite{Reinganum1900}, a paper, in which Drude's derivation of the
Wiedemann-Fanz law is reviewed and discussed with respect to
its underlying assumptions. To Mileva Mari\'c, he wrote: ``I found there a numerical
confirmation [...]  for the fundamental principles of the
electron theory, which filled me with real delight and completely convinced me 
about the electron theory.'' \cite[Doc.~111]{CPAE1}. Ten years later, Einstein 
expressed himself rather critical about Reinganum whose works he then characterized
as ``rather unclean'' (Einstein to Alfred Kleiner, 3 April 1912 \cite[Doc.~381]{CPAE5}).
See also Einstein to Hans Tanner, 24 April 1911 \cite[Doc.~265]{CPAE5} for another
critical comment on Reinganum's work and, for a general discussion of Einstein's 
early appreciation and later criticism of Drude's electron theory, see \cite{Renn2000}.}

The basic idea was to apply the concepts of the kinetic theory of
gases to a gas of electrons in the metal. Electrons were 
conceived of as particles with inertial mass and electric charge that
were moving about with random thermal motion in the metal. More
specifically, it was assumed that the electrons would not interfere or
interact with each other, and that they would only interact with the
positive ions upon collision. After colliding with an ion, an
electron would proceed on its path again freely, but with new energy and
momentum whose statistical distribution would only depend on the
place of the last collision.

The model allowed for a straighforward conceptualization of transport
phenomena such as heat conduction or electrical conduction. In order
to derive more specific relations for the quantities of interest,
further simplifications were usually made. Thus, in the beginning of
his course notes, Einstein sketched the derivation of a general
relation in the kinetic theory of gases that is applicable for generic
transport phenomena (``Transport of any Molecular Quantity through the
Gas.'') under the assumptions that all molecules at the same location
have the same mean velocity $c=\sqrt{\bar{c^2}}$. He considered a
molecular function $\mathcal{G}$ of an arbitrary quantity that is
being transported through the gas: 
\begin{quote} 
Each molecule carries along a certain quantity of something, with this
amount depending only on where the molecule's last collision took
place.\footnote{%
\cite[p.~183]{CPAE3}.}  
\end{quote}
And he computed the flux $\mathcal{F}$ of the molecular function
$\mathcal{G}$ by considering all molecules that contribute to the transport 
and by integrating over all directions. The result was
\begin{equation}
\mathcal{F} = - \frac{1}{3}nc\lambda \frac{\partial\mathcal{G}}{\partial z},
\label{eq:FGz}
\end{equation}
where $n$ is the number of molecules per unit volume, $\lambda$ the
mean free path, and the partial derivative is taken arbitrarily with
respect to the $z$-direction.

This relation is then quoted many pages later, when Einstein
set out to discuss the ``electron theory of m[etals].''\footnote{In
spite of differences in notation, Einstein's derivation closely
followed the one given in \cite{Drude1900}. There Drude, too, began by
quoting eq.~(\ref{eq:FGz}) from Boltzmann as his starting point.} He
first applied it to derive an expression for the thermal conductivity. Here the
molecular function is taken to be the kinetic energy of a ``molecule,'' i.e.\
an electron of mass $\mu$,
\begin{equation}
\mathcal{G} = \frac{1}{2}\mu c^2 = \frac{3RT}{2N},
\label{eq:G}
\end{equation}
which he relates to the temperature $T$ using the equipartition theorem. 
$R$ is the gas constant and $N$ is Avogadro's number. One has
\begin{equation}
\frac{\partial\mathcal{G}}{\partial z} = 
\frac{1}{2}\frac{\partial\mu c^2}{\partial T}\frac{\partial T}{\partial z}
= \frac{3R}{2N}\frac{\partial T}{\partial z},
\end{equation}
and hence
\begin{equation}
\mathcal{F} = -\frac{1}{2}\frac{R}{N}nc\lambda\frac{\partial T}{\partial z},
\end{equation}
from which one can readily read off the thermal conductivity $\kappa$ as the
(negative) coefficient in front of $\partial T/\partial z$,
\begin{equation}
\kappa = \frac{1}{2}\frac{R}{N}nc\lambda.
\label{eq:kappa}
\end{equation}
Note that the thermal conductivity still depends on 
the electron density $n$ and the mean free path $\lambda$
that are specific to individual metals.

The next step then is to obtain an expression for the electric
conductivity. Here the argument does not
go back to the general formula (\ref{eq:FGz}) of the flux for
the molecular function $\mathcal{G}$. Instead, Einstein's derivation started
from the concept of a mean collision time $\tau$, taken to be the
quotient of the mean free path and the mean velocity,
\begin{equation}
\frac{\lambda}{c} = \tau.
\end{equation}
In the absence of an external electric field $\mathcal{E}$, it is
assumed that the electrons are flying in different directions by equal
fractions and hence have no mean drift velocity $\mathcal{C}$. If,
however, an external electric field $\mathcal{E}$ is applied, it is
assumed that the electrons of charge $-\epsilon$ are accelerated during
the time of their free flight by a constant acceleration
$-\epsilon\mathcal{E}/\mu$. The mean drift velocity $\mathcal{C}$ was
then obtained by averaging over the mean free flight time as
\begin{equation}
\mathcal{C} 
= - \frac{1}{\tau}\int_{0}^{\tau}\frac{\epsilon\mathcal{E}}{\mu}t dt 
= - \frac{\mathcal{E}\epsilon}{\mu}\frac{\tau^2}{2}\cdot \frac{1}{\tau} 
= -\frac{1}{2}\mathcal{E}\frac{\epsilon}{\mu}\frac{\lambda}{c}.
\label{eq:averaging}
\end{equation}
Since a finite mean drift velocity $\mathcal{C}$ gives rise to a current density
$-n\mathcal{C}\epsilon$, one has
\begin{equation}
-n\mathcal{C}\epsilon = + \sigma\mathcal{E}
\end{equation}
and thus obtains the electric conductivity $\sigma$ as
\begin{equation}
\sigma = \frac{1}{2}\frac{\epsilon^2}{\mu}\frac{n\lambda}{c}.
\label{eq:sigma}
\end{equation}
Before discussing this expression let us complete the derivation by forming the
quotient of the thermal and electric conductivities to obtain the
Wiedemann-Franz law as
\begin{equation}
\frac{\kappa}{\sigma} = \frac{R}{N\epsilon^2}\mu c^2
= 3\frac{R^2}{N^2\epsilon^2}T.
\label{eq:kappaoversigma}
\end{equation}

The remarkable feature of this derivation of the Wiedemann-Franz law
is that it produces an expression for the Lorenz number $L$, i.e.\
the coefficient in front of $T$,
\begin{equation}
L \equiv \frac{\kappa}{\sigma T} = 3\frac{R^2}{N^2\epsilon^2},
\label{eq:Lorenz}
\end{equation}
that is in fairly good numerical agreement with the experimental
values.\footnote{For $R=8.31$~J/mol$\cdot$K, $N=6.02\times
10^{23}$/mol, $e=1.6\times 10^{-19}$C, we find $L$ to be $L\approx
2.2\times 10^{-8}$(J/molK)$^2$, a value which is within $10-20\%$ of
the experimentally observed value for many elements, see e.g.\
\cite[Tables I--VIII]{Meissner1920}, or
\cite[Table~1.6]{AshcroftMermin1976}.} 

Historically, this quantitative
agreement was of great significance, since it convinced most
theoreticians, including Einstein, that there was some truth to the
underlying model assumptions of the electron theory of metals. As it
turned out, however, this quantitative agreement is wholly
fortuitous. In our modern understanding of the issues at hand, it
arises from the cancellation of two factors of about one 
hundred.\footnote{See \cite[p.~23]{AshcroftMermin1976}.} The
electronic specific heat $c_v$ turns out to be a factor of $100$
smaller than the classical prediction $c_v=(3/2)nk_{\rm B}$, where
$k_B=R/N$. The mean square velocity of the electrons at room 
temperature, on the other hand, is about a factor of 100
larger.

Note, however, that already the numerical factor of $1/2$ in expression
(\ref{eq:kappa}) for $\kappa$, and hence also the numerical factor in 
the Wiedemann-Franz law (\ref{eq:kappaoversigma}), is an artifact, arising
(among other things) from the simplification that all molecules in the
same place would have the same mean velocity. A more careful
derivation of (\ref{eq:FGz}) would have to start from the full Maxwell
distribution, as was pointed out already by Drude
himself.\footnote{\cite[p.~569]{Drude1900}.} Such a refinement was 
carried out by Lorentz in 1905 who obtained a factor of $2$ 
instead of $3$ in the Wiedemann-Franz law 
(\ref{eq:kappaoversigma}). Other refinements of the derivation were also
discussed in the sequel and produced yet other numerical 
factors.\footnote{See, e.g., the discussion in \cite[pp.~785--791]{Seeliger1921}.}

One other problem needs to be mentioned here. It was pointed out, in a 
widely read modern textbook on solid state physics, that
eq.~(\ref{eq:kappaoversigma}), as it stands, is wrong by a factor of
$1/2$, since the electric conductivity $\sigma$ should actually be a
factor of two larger than that given in
eq.~(\ref{eq:sigma}).\footnote{See \cite[p.~23 and
prob.~1]{AshcroftMermin1976}. See also \cite[note 16]{Seeliger1921} and 
references cited therein for a contemporary discussion.} The claim here 
is that Drude's
erroneous result arises from an inconsistent application of the
underlying statistical assumptions.  The crucial point concerns the
assumptions about the statistical distribution of the times between
successive collisions. From a modern understanding, a natural assumption 
would be a Poissonian statistics, where the probability for any electron to undergo a
collision in the infinitesimal time interval $dt$ is proportional to
$dt/\tau^{\ast}$. Here $\tau^{\ast}$ is the mean collision
time, or more precisely the mean time between collisions in the trajectory of
a {\it single} electron. However, it also follows from the assumption of
a Poissonian statistics that the
mean time elapsed after the last collision for an electron {\it picked at
random} is also equal to $\tau^{\ast}$, as is the mean time until the next
collision of any such electron picked at random.\footnote{Roughly
speaking, the difference between the mean collision time of a single
electron and the mean free flight time of an electron picked at random
arises from the fact, that the probability distribution for the mean
free flight time is invoked twice in the computation of the latter case.}
Hence the mean time between successive collisions averaged over {\it all}
electrons is equal to $2\tau^{\ast}$. The averaging in
eq.~(\ref{eq:averaging}) should therefore be over
$(1/2\tau^{\ast})\int_0^{2\tau^{\ast}}$, or else by arguing that an electron picked
at random has been flying, on average, for a time $\tau^{\ast}$ thus
producing a mean drift velocity of $-\mathcal{E}\epsilon\tau^{\ast}/\mu$. A
similar error was not made, however, in the derivation of the thermal
conductivity (\ref{eq:kappa}). Hence, the theoretical account of the
Wiedemann-Franz law in eq.~(\ref{eq:kappaoversigma}) should have been
off by a factor of $2$ compared to the experimental data already on 
grounds of internal consistency of applying the model assumptions. 

Drude's result for the electric conductivity (\ref{eq:sigma}) is thus
incorrect if we assume a Poissonian statistics for the collisions of
the electrons in the metal. It is correct under the different and
rather restrictive assumption that the time $\tau$ between
collisions is always the same. In this case, and only in this case,
eq.~(\ref{eq:averaging}) still holds. Although the assumption of
a constant mean collision time was not made explicit in Drude's
original paper, it seems to me that it does not contradict any of his
explicit assumptions either, and the same holds for Einstein's
derivation in his kinetic theory lecture notes. After all, a similar
simplifying assumption was made quite explicitly about the mean
electronic velocity. Nevertheless, any non-trivial probability
distribution for $\tau$ would lead to numerical factors in
eq.~(\ref{eq:sigma}) that would be different from $1/2$, and that
would hence jeopardize the numerical agreement of the Lorenz number 
$L$ in eq.~(\ref{eq:Lorenz}) with the experimental data.

Drude's electron theory of metals thus had a curious epistemological
status. Its model assumptions were extremely simple and intuitive. It
allowed a more or less straightforward derivation of qualitatively
correct results about what quantities play a role in such phenomena as
electric conductivity. Some of these results turned out to be
completely independent of any microscopic details of the substance at
hand. The latter fact was in remarkable analogy to results in the
kinetic theory of gases, which had also quite successfully been able to
account for general regularities such as, e.g., the Dulong-Petit
law. Nevertheless, the quantitative, numerical results, although
in surprisingly good agreement with the available experimental data,
were somewhat fragile in the sense that modifications of the model or
of details of calculating the numerical results were not guaranteed
to maintain the agreement between theory and experiment.

With this general statement in mind, let us now comment more
specifically on the implications of expression (\ref{eq:sigma})
for the electric conductivity $\sigma$ in this model. For the purposes of our
present account, two things need to be pointed out. First, in contrast to
the Wiedemann-Franz law, the electric conductivity does depend on
material-specific quantities. Specifically, the result states that the
conductivity is proportional to the density $n$ of conduction
electrons and their mean free path $\lambda$. It was also seen to be
inversely proportional to the mean electronic velocity $c$, a quantity
that was naturally assumed to be only temperature
dependent. Other than that, the charge of the conduction electrons
$-\epsilon$ was a constant, as was their inertial mass $\mu$, as long as
relativistic effects were irrelevant.\footnote{Relativistic effects were, 
of course, irrelevant for Drude but recall that we are here discussing Einstein's
lecture notes of 1910 as background for a contemporary understanding of 
electric conduction.} The only quantities that
could therefore affect the temperature dependence of the conductivity
and account for its material specific features seemed to be the density of
conduction electrons and their free mean path. The historical
significance of this conclusion is illustrated in the following
comment that Einstein made in a letter to Lorentz written shortly after his
first visit to Leiden in 1911:
\begin{quote}
What I heard from Mr.~Kamerlingh Onnes and Mr.~Keesom was also very
important. It seems that the relationships between electrical
conductivity and temperature are becoming extremely important. If only
there would not always crop up the difficulty of one's not knowing
whether the change in the electrical conductivity should be attributed
mainly to the change in the number of the electrons or to the change
in their free path length, or to both. But I hope and am confident
that you will soon succeed in overcoming these
difficulties.\footnote{%
Einstein to Lorentz, 15 Feb 1911, \cite[Doc.~254]{CPAE5}.}
\end{quote}
Einstein had visited Leiden just a few weeks before
superconductivity was seen for the first time in the cryogenic
laboratory. His remark therefore reflects very precisely the
assumptions and expectations to which
theoretical physicists at the time, working as they did with a specific
model of an electron gas, would assimilate the discovery of a
sudden loss of resistivity.

A second comment on the significance of Drude's expression
(\ref{eq:sigma}) for the electrical conductivity follows from the
first. The experimental fact that in certain situations the
conductivity drops to an exceedingly small value, if not to $0$
altogether, immediately leads to a paradoxical situation when one tries
to assimilate the drop to the Drude model. In the case of the
phenomenological theory the defining relation of the electrical
conductivity (see eq.~(\ref{eq:jsigmaE}) below) degenerates for infinite
$\sigma$. In the model, too, the basic conceptualization of electrical
conductivity fails in such a limit: if, as seemed necessary, 
the temperature dependence of the
conductivity arises only from the number of available conduction
electrons and from their mean free path, it is immediately clear that
Drude's model cannot account for infinite conductivity. Given the
unambiguous experimental result that the loss of resistivity is at
least ten orders of magnitude compared to the resistance at room 
temperature, it is clear that with the sample sizes
at hand neither the number of free electrons nor the available space
for a large mean free path would permit an even roughly quantitative account 
of superconductivity.

One more comment may be in order before we proceed to discuss concrete
proposals of models for charge transport to account for
superconductive currents. While we are focusing for the purpose of
the present account on the theory of electrical conductivity, it
should be emphasized that the theoretical concepts and ideas that are
being invoked have more or less immediate implications for other
physical phenomena as well. The model's assumptions are accordingly constrained
by experimental knowledge that is directly relevant for other
consequences of the theory, such as specific heats, magnetic
properties, and the like. Conversely, the experimental fact of a superconductive state of
some metals at very low temperatures poses constraints on theoretical
considerations of other phenomena.  Einstein had had a long-standing interest in the
theory of specific heats, ever since his famous 1907 paper in which he
applied Planck's quantum hypothesis to the
problem.\footnote{\cite{Einstein1907a}.}
Because of the connections implied by the theoretical assumptions 
between different areas, it was natural for Einstein to
invoke the phenomenon of superconductivity in a consideration about
the existence of zero point energy:
\begin{quote}
There are serious doubts about the assumption of zero-point energy
existing in elastic oscillations. For if at falling temperatures the
(thermal) elastic vibrational energy does not drop to zero but only
{\it drops to a finite positive value}, then an analogous behavior
must be expected of all temperature-dependent properties of solids,
i.e., {\it the approach toward constant finite values} at low
temperatures. But this contradicts Kamerlingh Onnes's important
discovery, according to which pure metals become ``superconductors''
on approaching absolute zero.\footnote{%
The comment was published as a
discussion remark to Laue's presentation at the second Solvay Congress
\cite[p.~553]{CPAE4}, as a revised version of an original text that is
no longer available, see Einstein to Lorentz, 2 August 1915
\cite[Doc.~103]{CPAE8}.}
\end{quote}
The experimental discovery of superconductivity thus posed a challenge
to account for this phenomenon by modifying or substituting
model assumptions inherent in Drude's electron theory of
metals. Before proceeding to discuss these theoretical responses, we 
will briefly summarize what was known about superconductivity ca.~1920.

\section*{Superconductivity around 1920}
\addcontentsline{toc}{section}{Superconductivity around 1920}

By 1920, superconductivity was an anomalous and isolated, albeit
well-established phenomenon of cutting-edge technology.  It was in
Leiden that Kamerlingh Onnes had discovered the phenomenon in 1911,
three years after he succeeded in liquifying helium.\footnote{For
general historical accounts of the discovery and early developments in
the theory of superconductivity, see \cite{GG1989,Dahl1992,MV2003}.}
In fact, Onnes's cryogenic laboratory was the only laboratory in the
world able to achieve the liquefaction of helium at the
time. It retained this status until 1923, when the cryogenic
laboratory in Toronto liquified helium
with a copy of the Leiden cryogenic apparatus. In 1925, the low temperature
laboratory of the {\it Physikalisch-Technische Reichsanstalt} in Berlin
began to produce
liquid helium as well, and another such laboratory was established in
Charkov, Ukraine, in 1930.\footnote{\cite[p.~47]{MV2003}.}

Helium 
liquifies at atmospheric pressure at 4.22K. Since most metallic
superconductors have a transition temperature that is below the
boiling point of helium, it was only in Leiden that the phenomenon
could be, and was, found. It was observed first for mercury, which
has a transition temperature of 4.2K.
Measurements of the electrical resistance of mercury
at low temperatures were initially performed in order to
find a thermometric device for low temperatures,
that would replace thermometric measurements using the resistance of
platinum. Mercury, the only metal that is liquid at room temperatures,
was chosen because it was easiest to purify.

After establishing that the electrical resistance of mercury drops
very suddenly to a very low value at a certain temperature, the
phenomenon was further investigated. Around 1920, the following facts
about superconductivity had been established at the Leiden laboratory.%
\footnote{For contemporary reviews, see \cite{Crommelin1920}, 
\cite{Meissner1920}, \cite{Onnes1921b}.}

First of all, mercury was not the only substance that showed the
phenomenon. Four other metals were known in the early twenties to exhibit
superconductivity.\footnote{See, e.g., \cite{Onnes1924}.}  Tin
(Sn), was discovered to be superconducting in
1912,\footnote{\cite[p.~73]{Dahl1992}.} with a transition temperature
of 3.72K. Lead (Pb), which has a transition temperature of 7.19K, was
also found to be superconducting in 1912. However, here the precise
temperature of the transition was not explicitly observed or
determined because its transition temperature is in the temperature
range between the melting point of hydrogen at 13K and the boiling
point of helium where temperatures were not easily
determined.
Thallium (Tl) was discovered to be superconducting in 1919 with a
transition temperature of
2.32K.\footnote{ibid., p.~99--100.} In December 1922, indium
(In) was found to be superconductive at
3.41K.\footnote{ibid., p.~106.} However, gold (Au), iron
(Fe), platinum (Pt), cadmium (Cd), and copper (Cu) showed a finite and
constant electrical resistance at liquid helium temperatures.

As to the features of the superconductive transition, the following
facts had been established. The resistivity below the transition
temperature dropped to a value of order $10^{-10}$ as compared to that
at room temperatures. Upper limits on the residual resistance were
first determined by measuring potential drops along filaments carrying
large currents, later by the lifetime of persistent currents induced
in superconducting rings. The transition occurred within a narrow
temperature interval of the order of $10^{-3}$K. The superconducting
state was destroyed by critical currents of a certain value that
depended on the temperature. It was also destroyed by magnetic fields,
and it was determined that the threshold values were dependent on
the temperatures. The latter two features were thought to be
related, in that it was thought that the critical current is reached when
the induced magnetic field reaches a critical value.\footnote{This
hypothesis was known as the Silsbee
hypothesis, see \cite{Silsbee1916,Silsbee1917}.}

A controversial question concerned the influence of
impurities.  The drop of resistivity in Mercury seemed to be
independent of impurities, but the
purity of non-superconducting metals influenced the electric resistance at low
temperatures. The issue of impurities was a critical one, given
their significance in the theoretical account for electric resistivity in
the Drude model as well as in other models. In general, the role of
impurities remained an open issue due to difficulties in controlling and
determining the degree of purity. In particular, the available data did
not allow for an unambiguous decision as to whether ``really pure'' metals
like gold, iron, etc. would be superconducting at ``very low''
temperatures. As to the latter point, temperatures below 1.5K were 
very difficult to achieve, since the
vapor pressure of helium decreases rapidly with temperature. The low
temperature record was 0.8-0.9K and was attained by Onnes in
1921.\footnote{\cite[p.~133]{Dahl1992}.}

Many properties relating to
superconductivity had already been established before the
outbreak of World War I. During the war, low temperature research in
general and further research into the phenomenon itself was stalled, due both 
to shortage of personnel\footnote{The lack of personnel is mentioned
by Onnes who, himself almost 70 years of age, responds in a letter,
dated 13 August 1921, to Einstein's question about the empirical data
on the equations of state: ``We would have been further if only we had
more collaborators in order to undertake the numerous time-consuming
measurements that are necessary.'' 
(Albert Einstein Archives, The Hebrew University, (AEA) Call Number 
14-381). And at the end of that letter, Onnes asked Einstein
directly: ``How nice it would be if you could enthuse a well-trained
experimentor to come to Leiden in order to learn the determination of
equations of state at low temperatures and to continue these
investigations as a collaborator.'' 
} 
and of material resources, most importantly of sufficient supplies of
helium gas. But after the end of the war, low temperature research was
quickly resumed in Leiden with some significant experimental advances,
most notably an improvement of the cryogenic apparatus that allowed
the experimenter to physically remove the liquified helium from the
liquifier and transport it to experimental designs that no longer had
to be integrated with the liquifier.

\section*{Einstein's professional and personal ties to the Leiden
physicists}
\addcontentsline{toc}{section}{Einstein's professional and personal 
ties to the Leiden physicists}

Einstein was well informed about the work and experiments that were
being done
in Leiden.\footnote{For a discussion of Einstein's ties with Leiden,
see also \cite[pp.~xliii--xlviii]{CPAE10}.} As
early as 1901, the 22-year-old ETH graduate had sent a postcard to
Kamerlingh Onnes, who was looking for an assistant,
and applied for the position. Along with the postcard, Einstein sent an
offprint of his first published paper and a reply postcard which,
however, is still contained in the Kamerlingh Onnes
papers.\footnote{\cite{Proosdij1959} and \cite[Doc.~98]{CPAE1}.}

Ten years later, Einstein and Onnes exchanged offprints of their
respective recent publications, this time as colleagues, since
Einstein had recently been appointed associate professor at the
University of Zurich.\footnote{\cite[p.~623]{CPAE5}.} In his letter
to Kamerlingh Onnes, on 31 December 1910, sending his own
publications, Einstein also announces an imminent visit to Leiden:
\begin{quote}
In about a month's time I will have the extraordinary pleasure of
getting acquainted with you and your highly esteemed friend, Prof.\
Lorentz; for at that time I will deliver a lecture to the Leiden
Student Association.\footnote{%
Einstein to Kamerlingh Onnes, 31 December 1910, \cite[Doc.~242]{CPAE5}.}
\end{quote}
The lecture took place on 10 February 1911, and Einstein met
Kamerlingh Onnes just a few weeks before the discovery of
superconductivity.\footnote{For the chronology of the discovery, see
\cite[ch.~3]{Dahl1992}.} Apparently, the first encounter was
congenial. A few weeks later, after Einstein had accepted an offer at
the German University of Prague and had announced his
resignation from the
University of Zurich, Einstein and Kamerlingh Onnes corresponded about
Albert Perrier, a Swiss physicist then working as Onnes'
assistant who was being considered as Einstein's
successor.\footnote{See Einstein to Hans Schinz, 10 March 1911
\cite[Doc.~259]{CPAE5}.}

More important was the next encounter between Einstein and the Leiden
physicists at the first Solvay congress that took place from 27
October to 3 November 1911 in Brussels.\footnote{For a historical
discussion of the first Solvay congress, see \cite{Barkan1993}.} At
the meeting, Kamerlingh Onnes gave an account of the experiments
concerning electric conductivity at low
temperatures.\footnote{\cite{Onnes1912}.} His
participation at the first Solvay congress firmly established Einstein 
as a peer and congenial colleague of the Leiden physicists and, in fact, as
one of the 
leading theoretical physicists of the time. Just a few
weeks later, Einstein was asked for an opinion on the work of Keesom,
a student of Lorentz who was being considered for a vacant position in
Utrecht.\footnote{See Willem Julius to Einstein, 25 November and 29
December 1911 \cite[Docs.~314, 334]{CPAE5}.} And in a letter of 13
February 1912, Lorentz himself asked Einstein whether he would
consider becoming his successor in Leiden.\footnote{Hendrik A.\ Lorentz
to Einstein, 13 February 1912 \cite[Doc.~359]{CPAE5}. Einstein
declined immediately with some formal and polite excuses but added a
comment on his ``feeling of intellectual inferiority with regard to
you'' 
that may well have been the true reason for his decision: ``However, to
occupy your chair would be something inexpressibly oppressive for
me. I cannot analyze this in greater detail but I always felt sorry
for our colleague Hasen\"ohrl for having to occupy Boltzmann's
chair.'' (``Auf Ihrem Lehrstuhl zu sitzen, h\"atte etwas unsagbar
Dr\"uckendes f\"ur mich. Ich kann dies nicht weiter analysieren, aber
ich bemitleidete immer den Kollegen Hasen\"ohrl, dass er auf dem
Stuhle Boltzmann's sitzen muss.'' \cite[Doc.~360]{CPAE5}). In a letter
to Ehrenfest, Einstein even said that the offer ``had given him the creeps''
(``empfand ich ein unleugbares Gruseln''), Einstein to Ehrenfest, between 20 and 24 
December 1912 \cite[Doc.~425]{CPAE5}.} 
1912 is
also the year of the first personal encounter with Paul Ehrenfest,
who would instead become Lorentz's successor in Leiden,
when Ehrenfest visited Einstein in Prague. Ehrenfest
soon became one of Einstein's closest friends.\footnote{See
\cite[chap.~12]{Klein1970}  and also \cite{Einstein1934}.}

In August 1913, Einstein and Onnes met again when the
latter spent some time in a resort hotel in Baden
(Switzerland),\footnote{Einstein to Kamerlingh Onnes, 16 August 1913, 
and Kamerlingh Onnes to Einstein, 18 August 1913, 
\cite[Docs.~469, 471]{CPAE5}.}  and in March
1914 Einstein made another weeklong visit to Leiden on his way from
Zurich to Berlin.\footnote{\cite[p.~990]{CPAE8}.} By then he
was on a first-name basis with 
Ehrenfest\footnote{Einstein to Mileva Einstein-Mari\'c, 2 April 1914,
\cite[Doc.~1]{CPAE8}.} who in turn paid him another visit
in Berlin in May 1914.\footnote{\cite[p.~991]{CPAE8}.} During the war,
Einstein at first declined an invitation to visit Leiden in December 1915
because of family obligations\footnote{Einstein to Paul Ehrenfest, 26
December 1915, \cite[Doc.~173]{CPAE8}.} but then visited for
two weeks in late September and early
October 1916.\footnote{\cite[p.~1003]{CPAE8}.} When Ehrenfest invited him
again in late 1917, he was unable to come due to severe health problems
and the difficult travelling conditions:
\begin{quote}
[...] you can believe me that nothing is more appealing to me than a
trip to my dear Dutch friends, with whom I share such close and
kindred feelings in everything.\footnote{%
Einstein to Paul Ehrenfest, 12 November 1917,
\cite[Doc.~399]{CPAE8}.}
\end{quote}
Einstein's next visit to Leiden took place in October 1919, when
Einstein spent two weeks in the Netherlands where, among other things,
he attended a meeting of the Amsterdam Academy on 25 October 1919 in
which Lorentz informally announced results of the British eclipse
expedition. By this time, his Leiden colleagues had already been
trying to get Einstein to come to Leiden as a special
professor.\footnote{See Paul Ehrenfest to Einstein, 21 September
1919 and 24 November 1919 \cite[Docs.~109, 175]{CPAE9}. See also
\cite[pp.~xliii--xlviii]{CPAE10}.} He spent three weeks
in Leiden in May 1920, was inducted as foreign member into
the Royal Dutch Academy of Sciences on May 29, and also saw Onnes's 
laboratory:
\begin{quote}
Yesterday I visited Kamerlingh Onnes in his institute and attended a
nice lecture of his, saw interesting experiments.\footnote{%
Einstein to Elsa Einstein, 9 May 1920, \cite[Doc.~9]{CPAE10}.}
\end{quote}
A second trip that
same year took place in late October and early November, during which
he delivered his inaugural lecture and participated in the
meeting on magnetism mentioned above.

Except for Zurich, where he travelled frequently to see his
sons, Einstein visited no other place so frequently during
those years. We may thus assume that Einstein had regular, first-hand
information about what was going on in the Leiden cryogenic
laboratory.\footnote{Thirty years later, Einstein would remember his
relationship with Kamerlingh Onnes mainly as a personal friendship:
``I also knew Kamerlingh Onnes quite well but mainly
personally. Behind his warm and agreeable personality there was a
tenacity and energy that you only find very rarely. He was naturally
not so close to me in scientific matters, so that there were rarely
points for debate. Discussions with him were in general not easy since
he was extraordinarily precise in his intuitive thinking but could not
easily express himself clearly conceptually and was not easily
accessible to considerations of others, [...].'' 
Einstein to M.~Rooseboom, 27 February 1953 (AEA~14-396).}

We finally remark that a low temperature laboratory had also been
established in 1908 in the {\it Physikalisch-Technische Reichsanstalt} (PTR) in Berlin.
Einstein had been appointed member of the {\it Kuratorium} of the PTR in late 1916, 
regularly attended its annual meetings and actively participated in discussions
about its research.\footnote{See \cite{Hoffmann1980}.} 
Although not producing liquid
helium temperatures until 1925, the experimental and theoretical expertise of
his Berlin colleagues associated with this laboratory---Emil Warburg, Walther Nernst, 
Eduard Gr\"uneisen, Walther Mei{\ss}ner, and others---gave Einstein
further first-hand information about ongoing experimental research in the field
of low temperature physics.
%
%

\section*{Phenomenological theory of infinite conductivity}
\addcontentsline{toc}{section}{Phenomenological theory of infinite conductivity}

The fact that superconductors showed zero electric resistance was
experimentally well-established. Theoretically, this finding
posed a challenge since
the notion of infinite or perfect conductivity is conceptually
problematic. The concept of electrical conductivity is 
defined by the proportionality of current density $\vec{j}$ and
electric field $\vec{E}$,
\begin{equation}
\vec{j} = \sigma \vec{E}.
\label{eq:jsigmaE}
\end{equation}
To the extent that such a proportionality relation holds, the
constant $\sigma$ defines the electrical
conductivity. Setting aside complications such as anisotropies of the
conducting material, that render $\sigma$ a tensorial quantity,
frequency dependencies of $\sigma$ in the case of alternating
currents, or modifications of (\ref{eq:jsigmaE}) in the presence of
magnetic fields, the simple equation (\ref{eq:jsigmaE}) is
nevertheless constitutive of the very concept of conductivity. We see
immediately that this relation seems to lose all practical meaning in the limit of
$\sigma\rightarrow\infty$.

\subsection*{Lippmann's theorem}
\addcontentsline{toc}{subsection}{Lippmann's theorem}

The concept of infinite or perfect conductivity is nevertheless a
natural starting point for a theoretical analysis of the phenomenon of
superconductivity. Consequences of Maxwell's equations for 
metallic conductors of vanishing resisitivity were investigated
by Gabriel Lippmann (1845--1921) well before the discovery of
superconductivity.\footnote{For a brief discussion of Lippmann's
considerations of perfect conductivity, see
\cite[pp.~102--103]{Dahl1992}.} In 1889, Lippmann, professor of physics at
the Sorbonne who received the Nobel prize in 1908 for producing the first
color photographic plate, had published a short note in
the {\it Comptes rendus} on the law of induction in electric
circuits of vanishing resistance \cite{Lippmann1889}.

Contrary to what we have just said, in Lippmann's understanding, the
very notion of {\it finite} electrical conductivity was alien to the
fundamental laws of electrodynamics.\footnote{The very concept of electric
conduction was a problem for British field theoreticians but not so much
for the Continental tradition of electrodynamics. My discussion of Lippmann's
theorem is not meant as implying that the notion of electric resistivity, and more
specifically of vanishing electric resistivity, had not been 
a topic of theoretical discussion before. I discuss it here only as the most explicit 
discussion of the implications of superconductivity available for Einstein and
his contemporaries at the time. For general
accounts of the history of late nineteenth-century electrodynamics, see, e.g.,
\cite{Whittaker1951}, \cite{Whittaker1953}, \cite{Buchwald1985}, and 
\cite{Darrigol2000}. \label{note:disclaimer}} He compared the concept of
conductivity to the notion of friction in analytical mechanics,
where frictional forces are also not to be counted among the fundamental
concepts. For him it was hence rather natural to address
the case of perfect electrical conductivity if one wanted to come to an
understanding of the fundamental laws of electromagnetism.

Lippmann considered a conducting circuit where, in the absence of
external sources of voltage, the 
electromotive force $e$ is related to the electric current $i$ through
\begin{equation} 
e - L\frac{di}{dt} - ri = 0,
\label{eq:LippmannOhm}
\end{equation} 
with $L$ denoting the circuit's coefficient of self-induction and $r$ the resistance.

For such a conducting loop, the electromotive force $e$ is equal to
the change $dN$ in the number of magnetic flux lines per time due to external sources 
passing through the loop, 
\begin{equation} 
e = \frac{dN}{dt}.  
\end{equation} 

In addition, any induced currents will produce a change in the total
magnetic flux through the loop, 
\begin{equation} 
L\frac{di}{dt} = - \frac{dN'}{dt}, 
\end{equation} 
and so Ohm's law can be written as
\begin{equation} 
ri - \frac{dN}{dt} + \frac{dN'}{dt} = 0.
\end{equation} 
Setting now $r$ equal to $0$, one obtains after
integration,
\begin{equation} 
N + N' = \text{const}, 
\label{eq:NN0}
\end{equation} 
an equation that expresses the conservation of the magnetic flux
through the loop: 
\begin{quote} 
Put into words: {\it In a circuit devoid of any resistance, the
intensity of the induced current is always such that the magnetic flux
passing through the circuit remains constant.}\footnote{%
\cite[p.252]{Lippmann1889}.}  
\end{quote} 
In the remainder of his note, Lippmann then discussed implications of
equation (\ref{eq:NN0}) for superconducting coils and briefly observed
that an approximate analog of infinite conductivity would be given
experimentally for the rapidly oscillating Hertzian waves, pointing
to the fact that in this case the fields only penetrate into a small
surface layer of a metallic conductor.\footnote{Lippmann obviously
here refers to what is commonly known as the skin effect.}

In 1919, Lippmann took up his investigations of infinite conductivity
again, with explicit reference to Kamerlingh Onnes' discovery of
superconductivity.  In three only
slightly differing versions published in three different journals,\footnote{One
version, \cite{Lippmann1919a}, appeared in his own {\it Annales des
physiques}, a journal he was editing together with E.~Bouty, another
version, \cite{Lippmann1919b}, appeared in the {\it Comptes rendus}
(Lippmann being a member of the French Acad\'emie des Sciences since
1886), and a third version, \cite{Lippmann1919c}, was published in the
{\it Journal de physique th\'eorique et appliqu\'ee}.} he
referred to his earlier paper and its original motivation to
investigate electromagnetism without the friction-like concept of
finite electrical conductivity. He proudly pointed out that the
``fine experiments of Kamerlingh Onnes have brought about a physical
justification of the hypothesis of vanishing
resistance.''\footnote{%
\cite[p.~73]{Lippmann1919b}.}

Recapitulating the argument of his 1889 paper, Lippmann again considered
Ohm's law (\ref{eq:LippmannOhm}) which, for vanishing $r$, gives
\begin{equation} 
e = L\frac{di}{dt},
\label{eq:LippmannPC}
\end{equation} 
from which it follows immediately that a finite current density $i\neq
0$ can be maintained in the wire even in the absence of an
electromotive force $e$. Specializing to the case of a thin,
closed, homogeneous wire without any external sources of voltage or
soldered joints that might produce thermoelectric voltages, Lippmann
rederived his `th\'eor\`eme' of conservation of flux lines or, equivalently, of the
impenetrability for flux lines for an infinitely conductive ring. 
He noted that this theorem applies in particular to the
experiments on superconductors performed by Kamerlingh
Onnes. Curiously, Lippmann here cited nickel as a typical example of a
metal that loses its resistance at liquid helium
temperatures.\footnote{``A partir du moment o\'u le nickel est devenu
hyperconducteur, le nombre de lignes de force reste invariable.''
\cite[p.~248]{Lippmann1919a}. In \cite{Lippmann1919a}, but not in the
other two versions of his paper, Lippmann also mentions gold, along
side with lead, as one of the ``various metals'' 
whose resistance drops by a factor of at least $10^{10}$
\cite[p.~246]{Lippmann1919a}.}

While the previous argument held for loops of thin superconducting
wire, the general conclusions, argued Lippmann, remain true for
three-dimensional conductors such as a metallic bulk cylinder of
length $L$ and cross section $S$. A uniform magnetic field $H$
parallel to the axis of such a cylinder would penetrate the cylinder
in the case of finite conductivity, creating a magnetic moment of
size $SHL$. After cooling to the superconducting state, the flux would
remain frozen in, and the cylinder's magnetic moment would remain the
same. Similar conclusions would hold true for a hollow cylinder,
where the flux line distribution inside the cylinder would
remain the same, but one would find the lines slightly distorted.

From another point of view, the difference between perfect and finite
conductors could be interpreted as follows: In normal conductors, the
electromotive forces that induce the currents are proportional to the
relative velocity of field and conductor or to the temporal change of
an external magnetic field. In the case of perfect conductivity, on
the other hand, the electromotive forces only depend on the relative
displacement of field and conductor. The forces in the former case are
similar to viscous forces, while the electromotive forces in the
superconducting situation behave like elastic forces. They try to keep
the conductor at a fixed position which appears as a position of
equilibrium.

In the dynamical case of electromagnetic waves, Lippmann repeated his
observation about the known fact that electromagnetic waves do not
penetrate into the bulk of metallic conductors of high
conductivity. This behavior, he remarked, carries over to the
superconducting case. Here again, electromagnetic waves cannot
penetrate into the superconducting bulk substance.

Lippmann concluded his note with comments on the peculiarities of the
transmission of forces between two superconducting rings, and 
on interpreting Amp\`erian molecular currents in terms of
perfect conductivity.

\subsection*{Ehrenfest's diaries}
\addcontentsline{toc}{subsection}{Ehrenfest's diaries}

We have no direct evidence that Einstein was aware of Lippmann's
papers, but we do have some indirect evidence that he knew about
Lippmann's considerations.  Lippmann's name is mentioned in
Ehrenfest's diaries in an entry ``Supraleiter-Hall-Effect (Lippmann)''
found next to other entries dated April
1920.\footnote{Ehrenfest Archive, Museum Boerhaave (Rijksmuseum voor
de Geschiedenis van de Natuurwetenschappen en van de Geneeskunde),
Leiden, Notebooks, ENB:1-26/2.}  Since Einstein visited Leiden
from 7 to 27 May 1920,\footnote{See \cite[pp.~570, 572]{CPAE10} and
Einstein to Elsa Einstein, 27 May 1920 \cite[Doc.~32]{CPAE10}.}  we
may assume that the topic was discussed by Ehrenfest and
Einstein during that visit.\footnote{Einstein's name is mentioned
frequently in Ehrenfest's diaries, as are mentions of the problem of
superconductivity: Entry 5463, following an entry explicitly dated to
31 May 1920, reads: ``Precessionsbeweg. von Str\"omen in
Supra-leitern - Kugel wegen a.)  Tr\"agheit der Elektronen b.)
Hall-Effect.'' (ENB 1-26/6); entry 5470, found between entries dated
14 June 1920 and 2 July 1920, again says: ``Supraleiter mit
Hall-Effect.'' (ENB 1-26/7). See also the discussion of the
``Magnet-Woche'' below.}

\begin{figure}[thb]
\begin{center}
\includegraphics[scale=0.50]{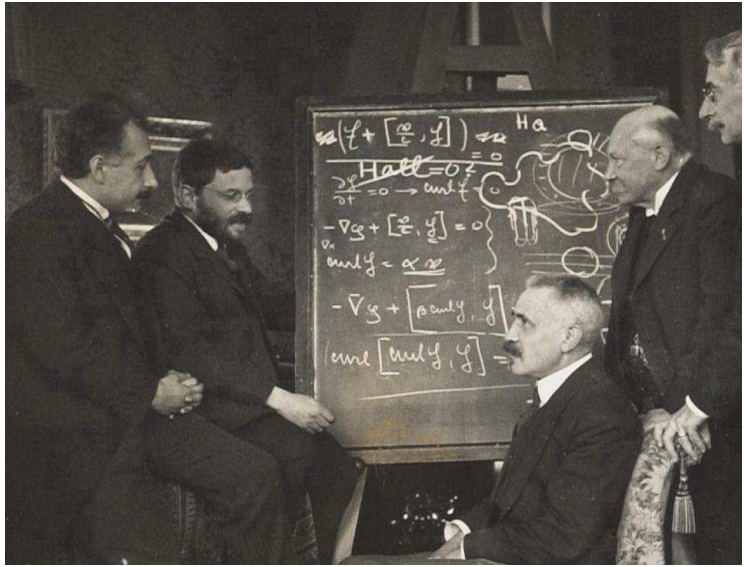}
\caption{Albert Einstein, Paul Ehrenfest, Paul Langevin, Heike Kamerlingh 
Onnes, and Pierre Weiss discussing superconductivity during the 
``Magnet-Woche'' in Leiden in November 1920 (Photo: AIP).}
\label{fig:blackboard}
\end{center}
\end{figure}

Indeed, Ehrenfest's diaries contain a more elaborate entry on this
topic. Entry 5548 is found next to an entry that describes Einstein's
visit to Leiden in November.\footnote{The diary entry on Einstein's
visit reads: ``Magnet-Woche: Einstein allein Ankunft am Abend[---]
kleines Fenster [---] alle jubelnd hinaus. Wandert zu Onnes ||
Spaziergang Haagsche Weg Goldnebel (Ruhe, Weide, Kirchhoff)
Triospielen bei Maler Onnes. Ein Abendessen in grossem dunklen
Esszimmer Einstein mit [--] Langevin rauchend auf eiskalter
Nachtstrasse Weiss Langevin, Lorentz, Einstein, Taniz,
Woltjer-Sonne.''} The entry itself is then dated 2 November and
entitled ``Hall-Effect im Supraleiter.'' The consideration and
equations of this entry actually closely parallel those found
on the blackboard on a photograph taken, in all probability, during the
``Magnet-Woche'' and showing (from left to right) Einstein,
Ehrenfest, Langevin, Kamerlingh Onnes and Pierre
Weiss, see Fig.~\ref{fig:blackboard}.\footnote{%
The photo is also shown on the jacket cover of \cite{CPAE10}.} 
The equations on the blackboard appear to be written by
Ehrenfest, who also poses in the photograph as if he were the
one writing on the blackboard. Let us briefly review the consideration in
Ehrenfest's diary with cross-reference to the blackboard image.

Ehrenfest began by writing down the following condition
\footnote{In the following, I will translate the equations as they appear
in Ehrenfest's and Einstein's manuscripts into a unified
notation, substituting, e.g., $\vec{E}$ for $\mathcal{E}$, and expressing
vector analytic expressions throughout in terms of the Nabla-operator
$\vec{\nabla}=(\partial_x, \partial_y, \partial_z)$.}
\begin{equation}
\vec{E} + \alpha\frac{\vec{v}}{c}\times \vec{H} = 0.
\label{eq:EhrenfestPC}
\end{equation}
Here $\vec{E}$ and $\vec{H}$ denote the electric and magnetic
field vectors, $\vec{v}$ the local velocity of the current
carrying charges, $c$ the speed of light (using c.g.s
units), and $\alpha$ is a numerical parameter which is included only
in the diary version and whose significance will become clearer below.
The square brackets denote the vectorial cross product. Equation
(\ref{eq:EhrenfestPC}) captures the condition of perfect conductivity.
Since it will be the basis for much of the following let us discuss
its significance in some more detail. Consider first a resistor for which
Ohm's law holds between the voltage $U$ and the current $I$
in its integral form,
\begin{equation}
U = R\cdot I.
\end{equation}
Here $R$ is the
total electrical resistance of a piece of current carrying matter of,
say, cylindrical shape with length $L$ and cross-section $S$. If we
assume homogeneity along the cylinder, we can relate the voltage drop
$U$ to the local electric field strength $E$ along the wire as
$U=EL$, the total resistance $R$ to a local resistivity $\rho$ as
$R=\rho L/S$, and the total current $I$ to a local current density $j$ as
$I=jS$. We thus obtain a local version of Ohm's law, 
\begin{equation}
E = \rho j,
\label{eq:Erhoj}
\end{equation}
that is independent of the geometric shape of the resistor.  The
latter equation turns into equation (\ref{eq:jsigmaE}) if we identify
the local resistivity $\rho$ as the reciprocal of the conductivity,
\begin{equation}
\sigma=1/\rho,
\end{equation}
and take into account the vector character of the
current density and the electric field. Such a distinction between an integral and 
a local version of Ohm's law was standard textbook knowledge of the time, 
as witnessed, e.g., in \cite[\S~53.]{Foeppl1907} where the integral version
is said to reflect directly an empirical fact whereas the differential law would be
more suitable for theoretical analysis. Recalling now that the Lorentz
force expression reads
\begin{equation}
\vec{F} = \rho_{\rm e}\left(\vec{E}+\frac{\vec{v}}{c}\times\vec{H}\right),
\end{equation}
where $\vec{F}$ is the force density and $\rho_{\rm e}$ the electrical
charge density moving with velocity $\vec{v}$, 
it is natural to add a
term proportional to the cross-product of the velocity $\vec{v}$ of
the charge carriers and the magnetic field $\vec{H}$, to obtain a
generalized and local version of Ohm's law in the form
\begin{equation}
\vec{j} = \sigma\left(\vec{E} + \alpha\frac{\vec{v}}{c}\times\vec{H}\right),
\end{equation}
where we have again introduced an arbitrary numerical factor $\alpha$.
In this version of Ohm's law, one can now take again the limit of infinite
conductivity $\sigma\rightarrow\infty$ in a sensible way to obtain
Ehrenfest's condition of infinite conductivity in the form of equation 
(\ref{eq:EhrenfestPC}).

Lippmann's condition of
perfect conductivity (\ref{eq:LippmannPC}) was obtained in a somewhat
analogous manner from Ohm's law (\ref{eq:LippmannOhm}) for vanishing
resistance $r$. However, Lippmann worked with the total current in a
circuit rather than a local version of Ohm's law valid at any point within a 
conductor. Hence, his induction term $Ldi/dt$ is different from the
magnetic term $\vec{v}\times\vec{H}$ in Ehrenfest's version.

Equation (\ref{eq:EhrenfestPC}) is to be investigated in order to
understand infinite conductivity. Ehrenfest does so by invoking
Amp\'ere's law,
\begin{equation}
\beta\vec{v} = \vec{\nabla}\times\vec{H},
\label{eq:Ampere}
\end{equation}
where $\beta\vec{v}=(4\pi/c)\rho_{\rm e}\vec{v}$ would be the current density,\footnote{On
the blackboard, Ehrenfest used $\alpha$ instead of $\beta$.} and any
displacement current terms are taken to be negligible. Amp\'ere's law allows him to 
eliminate $\vec{v}$ from the condition of infinite conductivity. 
He also assumes that the magnetic
field does not change with time,\footnote{The time independence of
$\vec{H}$ is written down as a condition explicitly on the
blackboard.} which, by virtue of Faraday's law, implies that the
electric field is irrotational and hence has a potential $\varphi$ as
\begin{equation}
\vec{E} = - \vec{\nabla} \varphi.
\label{eq:Ephi}
\end{equation}
Using (\ref{eq:Ampere}) and (\ref{eq:Ephi}), we can hence write 
(\ref{eq:EhrenfestPC}) as
\begin{equation}
-\vec{\nabla}{\varphi} + \frac{\alpha}{\beta c} (\vec{\nabla}\times\vec{H}) \times\vec{H} = 0.%
\footnote{This equation is not written down in the diary, but it is written on the
blackboard with a different notation for the constant in front of the second term.}
\label{eq:curlcurlphi}
\end{equation}
In a final step, Ehrenfest now takes the rotation of (\ref{eq:curlcurlphi}) and 
obtains the condition
\begin{equation}
\vec{\nabla}\times\left[(\vec{\nabla}\times\vec{H})\times\vec{H}\right] = 0,
\label{eq:curlcurl}
\end{equation}
as a characteristic condition for the magnetic field in
superconductors in time-independent situations.

In order to see the consequences of (\ref{eq:curlcurl}), Ehrenfest 
rewrote it, using standard equations of vector calculus, more
explicitly as
\begin{equation}
[(\vec{\nabla}\times\vec{H})\cdot\vec{\nabla}]\cdot\vec{H}
- (\vec{H}\cdot\vec{\nabla})(\vec{\nabla}\times\vec{H})
+ \vec{H}\cdot[\vec{\nabla}\cdot(\vec{\nabla}\times\vec{H})]
- (\vec{\nabla}\times\vec{H})\cdot(\vec{\nabla}\cdot\vec{H}) = 0.
\end{equation}
The third term vanishes identically because it is the divergence of a
rotation, and the fourth term vanishes on account of Maxwell's
equations.
The remaining first two terms were then written out explicitly as
\begin{align}
&\left[\left(\frac{\partial H_z}{\partial y}
-\frac{\partial H_y}{\partial z}\right)\frac{\partial}{\partial x}
+ \left(\frac{\partial H_x}{\partial z}
-\frac{\partial H_z}{\partial x}\right)\frac{\partial}{\partial y}
+\left(\frac{\partial H_y}{\partial x}
-\frac{\partial H_x}{\partial y}\right)\frac{\partial}{\partial z}
\right]\cdot\notag \\
&\cdot\left(\mathfrak{i}H_x + \mathfrak{j}H_y + \mathfrak{k}H_z\right)
-\notag\\
&-\left(H_x\frac{\partial}{\partial x} + H_y\frac{\partial}{\partial y} 
+ H_z\frac{\partial}{\partial z}\right)
\left\{\mathfrak{i}\left(\frac{\partial H_z}{\partial z}
-\frac{\partial H_y}{\partial z}\right)\cdots + 
\cdots\right\} = 0.
\label{eq:discutieren}
\end{align}
with orthogonal unit vectors $\mathfrak{i}$, $\mathfrak{j}$, $\mathfrak{k}$.
At this point in his diary, Ehrenfest adds the comment ``discuss!''
(``discutieren!''), and obviously this is also what Einstein,
Ehrenfest, Langevin, Onnes, and Weiss are posing to do on their
photograph. But instead of discussing (\ref{eq:discutieren}) any 
further,\footnote{On
an earlier but closely related page (ENB 1-26/6), Ehrenfest is also
concerned with a discussion of (\ref{eq:curlcurl}) but again does
not proceed any further than by looking at components of
(\ref{eq:curlcurl}) written out explicitly.} Ehrenfest goes back to
the original expression for perfect conductivity (\ref{eq:EhrenfestPC})
and rewrites it, using (\ref{eq:Ephi}) as well as 
$\vec{j}\equiv\rho_{\rm e}\vec{v}$ in the form
\begin{equation}
-\vec{\nabla}\varphi + \frac{\alpha}{\rho_{\rm e}}\frac{\vec{j}}{c}\times\vec{H} = 0.
\end{equation}
He immediately concludes that it follows that the electrostatic
potential $\varphi$ is constant along lines parallel to either the
current density $\vec{j}$ or the magnetic field lines $\vec{H}$.

A discussion of expression
(\ref{eq:discutieren}), e.g. by specializing to certain symmetries,
fields, etc., would be the natural thing to do, and, in fact,
Ehrenfest began to simplify
(\ref{eq:discutieren}) for the case where all derivatives with respect
to $z$ would vanish. But this calculation breaks off.
Indeed, an exploration of equation (\ref{eq:discutieren}) or even
of (\ref{eq:curlcurl}) would not be too enlightening in the end since
all time dependence had been assumed absent from the outset anyway.

The calculation on the Hall effect in superconductors in Ehrenfest's diary 
proceeded on the basis of the classical Maxwell equations and 
explored the implications of perfect conductivity. The latter condition
was expressed in terms of equation (\ref{eq:EhrenfestPC}). In his
first calculation, Ehrenfest deduced from this {\it ansatz} a vector
differential equation for the magnetic field (\ref{eq:curlcurl}) that
does not contain any sources or currents. The equation was not explored
any further, and it is unclear what
conclusion Ehrenfest may have drawn at this point. However, we do have
indications that these issues were further pursued in the
discussions between Ehrenfest and Einstein and possibly other
participants of the ``Magnet-Woche.''

\subsection*{Theorizing about experiments on the Hall effect for superconductors}
\addcontentsline{toc}{subsection}{Theorizing about experiments on the Hall
effect for superconductors}

Einstein, too, thought along this line of
exploring consequences of Maxwell's equations for infinite
conductivity. Three different and independent
sources all document the very same consideration. One 
source is another entry in Ehrenfest's diary, in which he
excerpted an argument from a (non-extant) letter by Einstein, dated 9 December 
1920.\footnote{``from letter by Einstein 9 XII 1920.'' ENB 1-26/46 and
1-26/47.  The entry is numbered as 5559 but this is actually the
second number with this entry since on the previous page, ENB 1-26/45,
Ehrenfest had already recorded (unrelated) entries 5559 to 5564. Quite
possibly Ehrenfest had opened his notebook on p.44 which ends with an
entry 5558, then mistakenly turned over two pages at once and
continued on p.~46 with another entry 5559. The entry with Ehrenfest's
excerpt is published as \cite[Doc.~227]{CPAE10}.}
Calculations by Einstein along the same line are also found on a single 
manuscript page, dated in an unknown hand to 12 December 1920, located
at the Burndy library.\footnote{The manuscript page is 
extant in the Burndy Library,
Cambridge, Ma. A pencil note on the back reads: ``Manuskript und
Zeichnungen von Prof. Albert Einstein 12 XII 20.'' The manuscript is
published as \cite[Appendix]{CPAE10}. I wish to thank P.~Cronenwett 
for providing the Einstein Papers Project with
high-quality scans of the Burndy library manuscript.}  
And the very same argument is
finally also laid out in a letter by Einstein to Lorentz, dated 1
January 1921.\footnote{AEA~16 533.}

I will here give a presentation of the argument that is not literally faithful
to the originals but is in itself complete and notationally
consistent. Special features of the individual source documents will
be pointed out along the way.\footnote{%
The three sources differ among each
other in notation, in the degree to which the relevant equations were
written out and commented on, as well as in the existence of
illustrative figures. None of the three sources present the
argument more comprehensively than any of the other.
The Burndy manuscript is a little more complete in the equations that
Einstein actually wrote down but Ehrenfest's letter excerpt and Einstein's 
letter to Lorentz are more explicit about the meaning of the calculations.}

\begin{figure}[thb]
\begin{center}
\includegraphics[scale=.35]{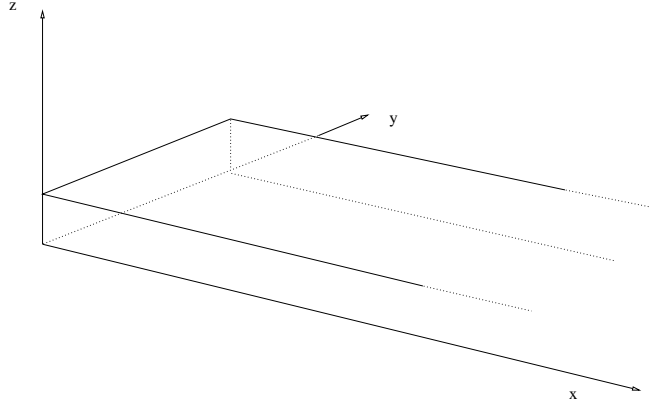}
\caption{A slab of superconducting metal extending along the $x$-axis.}
\label{fig:slab}
\end{center}
\end{figure}

Einstein works out on consequences of the 
condition for perfect conductivity (\ref{eq:EhrenfestPC}), which we will rewrite
here in the form
\begin{equation}
\vec{E} = -\tilde{\alpha}\frac{\vec{j}}{c}\times\vec{H}
\label{eq:EinsteinPC}
\end{equation}
for electric and magnetic fields $\vec{E}$ and $\vec{H}$ and current density
$\vec{j}$. $\alpha=\tilde{\alpha}\rho_{\rm e}$ is again a numerical parameter 
to be discussed below.
Instead of invoking Amp\'ere's law at this point (see (\ref{eq:Ampere}) above), 
as Ehrenfest had done, Einstein started from Faraday's law of induction
\begin{equation}
\vec{\nabla}\times\vec{E} + \frac{1}{c}\partial_tH = 0.
\label{eq:Faraday}
\end{equation}
Taking the rotation of (\ref{eq:EinsteinPC}) and substituting
$-\frac{1}{c}\partial_t\vec{H}$ for $\vec{\nabla}\times\vec{E}$ then yields
\begin{equation}
\tilde{\alpha}\vec{\nabla}\times(\vec{j}\times\vec{H}) + \partial_t \vec{H} = 0.
\label{eq:curlBdt}
\end{equation}
which, written explicitly in components, reads
\begin{align}
\tilde{\alpha}\partial_y\left(j_xH_y-j_yH_x\right) 
- \tilde{\alpha}\partial_z\left(j_zH_x-j_xH_z\right) + \partial_tH_x &= 0,
\label{eq:x_comp} \\
\tilde{\alpha}\partial_z\left(j_yH_z-j_zH_y\right) 
- \tilde{\alpha}\partial_x\left(j_xH_y-j_yH_x\right) + \partial_tH_y &= 0, \\
\tilde{\alpha}\partial_x\left(j_zH_x-j_xH_z\right) 
- \tilde{\alpha}\partial_y\left(j_yH_z-j_zH_y\right) + \partial_tH_z &= 0. 
\label{eq:z_comp}
\end{align}
Einstein's consideration now rests on an interpretation of the
characteristic equation (\ref{eq:curlBdt})
viz. (\ref{eq:x_comp})--(\ref{eq:z_comp}) just as Ehrenfest was
trying to interpret his (\ref{eq:curlcurl}). In his
letter to Lorentz, Einstein wrote eqs.~(\ref{eq:EinsteinPC}),
(\ref{eq:Faraday}), and (\ref{eq:curlBdt}), and continued ``to
investigate the evolution of the phenomenon in a slab that carries a
current in the $x$-direction.''
For illustration, Einstein included a sketch of a thin superconducting 
slab, see Fig.~\ref{fig:slab}.

He first assumed that there is no component of the current in
the vertical or $z$-direction,
\begin{equation}
j_z \equiv 0,
\label{eq:cond1}
\end{equation}
and that only $j_x$ and $j_y$ have non-vanishing values. Second, he
assumed that the magnetic field only has a vertical component $H_z$,
and the $x$- and $y$-components vanish,
\begin{equation}
H_x \equiv H_y \equiv 0.
\label{eq:cond2}
\end{equation}
Third, if moreover none of the fields and quantities (in the slab)
change along the $z$-direction, i.e. 
\begin{equation}
\partial_z\equiv 0,
\label{eq:cond3}
\end{equation}
eqs.~(\ref{eq:x_comp})--(\ref{eq:z_comp}) reduce to 
\begin{equation}
-\tilde{\alpha}\partial_x(j_xH_z) - \tilde{\alpha}\partial_y(j_yH_z) + \partial_tH_z = 0.
\label{eq:rotBdtred}
\end{equation}

Conditions (\ref{eq:cond1})--(\ref{eq:cond3}) and the specification to
a flat current carrying slab are very suggestive of an
experiment designed to investigate the transverse Hall
effect.\footnote{
For a historical discussion of the Hall effect, see
\cite[Part~II and App.~3]{Buchwald1985}. For a contemporary
discussion, see, e.g., \cite{Beckman1922}.} And this is, in fact, what
Einstein here had in mind. He continued by stating that
the $y$-components of the current are ``induced by the Hall
effect'',
and that one may assume, with good approximation, that
\begin{equation}
\partial_y (j_yH_z) \equiv 0.
\label{eq:cond4}
\end{equation}
With this assumption, eq.~(\ref{eq:rotBdtred}) further simplifies and
its solutions are of the form (due to the continuity equation and since the slab
is assumed to be flat, we have $\partial_xj_x=0$) 
\begin{equation}
H_z = f(x-\tilde{\alpha} j_x t),
\end{equation}
for some arbitrary function $f$ which he interpreted as follows: 
\begin{quote} 
The magnetic field is hence dragged along by the current with 
velocity $\tilde{\alpha} j_x$.
\end{quote}
This consequence is, in fact, a general property of the
condition of perfect conductivity. In ideal magnetohydrodynamics,
e.g., it is shown on similar grounds that the magnetic field lines 
move along with the current in an ionized plasma.\footnote{See the
discussion below on p.~\pageref{plasma}.}  We now also see the
significance of the numerical parameter $\alpha$.  If $\alpha$ is
smaller than $1$, the field lines are being ``dragged along'' 
with a velocity that is reduced by a factor of $\alpha$ compared to
the velocity of the current.

Einstein continued in his letter to Lorentz:

\begin{quote}

For a discontinuous change of the slab's thickness, [$j_xH_z$] is
continuous or also [$\frac{1}{\delta}\cdot
H_z$], where $\delta$ is the slab's thickness.
\end{quote}
He concluded by suggesting an experimenal investigation:
\begin{quote}
Thus we sufficiently understand what processes to expect in order to
be able to decide experimentally whether the Hall effect exists at low
temperatures.
\end{quote}

\begin{figure}[thb]
\begin{center}
\includegraphics[scale=.35]{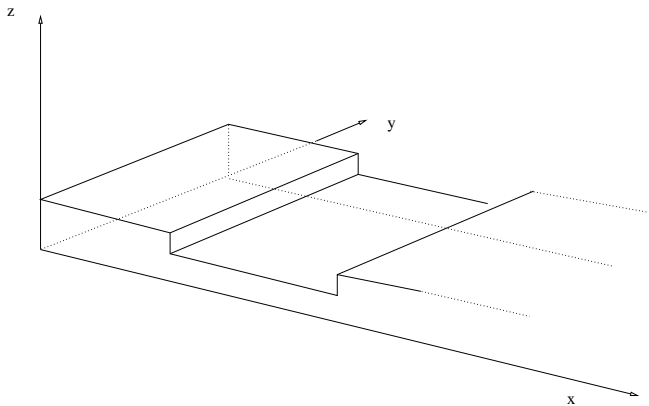}
\caption{A slab of superconducting metal of varying thickness 
extending along the $x$-axis.}
\label{fig:slab2}
\end{center}
\end{figure}

For the last step, Einstein clearly assumed that the term $\partial_tH_z$
vanishes, which leaves us with
\begin{equation}
j_xH_z=\textrm{const}.
\label{eq:jBconst}
\end{equation}
Consider then a slight variant of the slab, like the one in 
fig.~\ref{fig:slab2} where the thickness varies along the $x$-direction.
Since we would naturally assume charge conservation,
\begin{equation}
\vec{\nabla}\times\vec{j}\equiv 0,
\end{equation}
for the superconducting current, the $x$-component of the current
would vary in proportion to the thickness $\delta$. In order to satisfy
(\ref{eq:jBconst}), the magnetic field component $H_z$ would therefore have to
vary in inverse proportion to the thickness $\delta$, as stated by Einstein.
Einstein seems accordingly 
that the transverse Hall voltage along a slab of varying
thickness $\delta$ should vary inversely as $\delta$, and that this hypothesis
should be put to experimental test.

With slight variations, Ehrenfest's excerpt notes present Einstein's
argument in a similar manner to our presentation above. But
his discussion of (\ref{eq:jBconst}) is a little different from
the one that Einstein gave in his letter to Lorentz. With reference 
to the situation of a slab of varying
thickness as depicted in fig.~\ref{fig:slab2}, Ehrenfest argued as follows.  
Let the magnetic field $H_z$ be constant at some initial point
at time $t=0$. If one now turns on the current,
\begin{quote}
the current thus creates at its onset at first a point $A$ (or $B$
for neg[ative] $\tilde{\alpha}$) where the field is smaller, which point then
runs with velocity [$\tilde{\alpha} j_x$] along the thin part of the slab. The
field in the thick part remains permanently constant.\footnote{%
\cite[Doc.~227]{CPAE10}.}
\end{quote}
The Burndy manuscript version of
the argument is less explicit and more sketchy. It also shows a few
fragmentary equations involving the current four-vector, the electromagnetic
field tensor and a stress-energy tensor in
four-dimensional, Lorentz covariant notation.

I have not found any indication that Einstein's argument was discussed
anywhere in print, nor did I find any indication that the hypothesis
of a varying magnetic field in superconducting slabs whose
thickness changes from point to point was ever tested directly 
and explicitly experimentally. It
seems likely that the technological possibilities of the Leiden
cryogenic laboratory at the time cwwere inadequate to
produce superconducting slabs of varying and controllable thickness and
to measure a Hall voltage with sufficient spatial
resolution. In the concluding section of his 1921 Solvay report on
superconductivity, Kamerlingh Onnes points out that the investigation
of the phenomenon of superconductivity is complicated enough without
external fields:
\begin{quote}
By introducing an external field, every question is doubled, as it
were. Others are added. We would enter here into a vast terrain, where
almost all experimental investigations are wanting.\footnote{%
\cite[p.~50]{Onnes1921b}.}
\end{quote}
And referring back to pre-war experiments on the Hall effect
\cite{OnnesHof1914}, Onnes continues
\begin{quote}
It is only the Hall phenomenon on which investigations have been
made, which have shown that the electromotive force that is observed
in the usual way disappears with the resistance as soon as
superconductivity appears.\footnote{%
ibid., p.~50. For an account of the Leiden experiments on the Hall effect at
low temperatures, see also \cite{Beckman1922}.} 
\end{quote}
The experiments referred to had been done in order to investigate the
influence of a magnetic field on the electric conductivity. Onnes and Hof 
had investigated plates of tin and lead and found that a Hall effect was 
observed at liquid helium temperatures for magnetic fields that were high 
enough to destroy the superconductivity. But for low magnetic fields the 
Hall voltage was found to vanish just as did the electrical resistance. It is
clear that those experiments were not sophisticated enough to provide the kind
of spatial and temporal resolution that Einstein's idea would require.

The situation might have changed in the late twenties or early thirties
with other cryogenic laboratories capable of investigating superconductivity
entering the scene. But then again,
an experiment such as the one envisaged here would not have made much 
sense after the discovery of the Meissner effect in 1933. Once it was
realized that superconductivity is a thermodynamic state
characterized not only by infinite conductivity, but also by perfect
diamagnetism, it would have become clear that
the magnetic field would be expelled from the superconducting slab
rather than be dragged along with the current flowing inside
it.\footnote{In actual experiments of the kind suggested by Einstein, 
other effects may play a role, too, e.g.\ intermediate states of only partially 
expelled magnetic fields, see \cite{Huebener2001} for a discussion of 
magnetic flux effects in superconductors. Note also that since the 
Meissner effect concerns only bulk properties, Lippmann's theorem of 
conservation of magnetic flux through a looped circuit still holds good.}

After 1933, phenomenological theories 
of superconductivity also needed to account for perfect
diamagnetism. This task was successfully achieved in 1935 through a
modification of Maxwell's equations proposed by the
brothers Fritz and Heinz London. 
In this theory, the current $\vec{j}$ is supposed to consist
of two components, a normal component $\vec{j}_n$ and a
superconducting component $\vec{j}_s$. For the superconducting
component, one still has infinite conductivity $\sigma_s=\infty$ but
for the normal component one has a modification of the Maxwell
equations, given by the so-called London equations,
\begin{equation}
\lambda\partial_t\vec{j}_n = \vec{E},
\end{equation}
and
\begin{equation}
\lambda\vec{\nabla}\times\vec{j} = - \vec{H}.
\end{equation}
On the basis of these equations, it can be shown that magnetic fields
may only penetrate into the superconductive bulk matter up to
a distance of order $\lambda$.\footnote{For historical discussion 
of the London theory of superconductivity, see \cite{Gavroglu1995} and 
\cite[chap.~11]{Dahl1992}.}

\label{plasma}
The condition of perfect conductivity (\ref{eq:EhrenfestPC}) or
(\ref{eq:EinsteinPC}) which was at the core of the phenomenological
theory described in this section thus was no longer valid in
the theory of superconductivity after 1933. Yet, the
condition of infinite conductivity still plays a role today
in the context of plasma physics, more speficically in the conceptual
framework of ideal magnetohydrodynamics. The theory of an ionized plasma
at low frequencies is again given by Maxwell's equations plus the
condition of infinite conductivity. Indeed, the general conclusion of
Lippmann and Einstein of a freezing in of the magnetic flux lines
carries over to the case of a magnetohydrodynamic liquid. The
difference here is that the positive ions now also come into play,
leading to the possibility of an energy transfer between
electromagnetic field energy and kinetic energy of the positive
ions. Adding an equation of motion for a charged liquid with mass
density of the distributed positive ions, Hannes Alfv\'en first showed the
possibility of the existence of so-called magnetohydrodynamic waves.
In these waves, magnetic flux lines perform an oscillatory motion with the
charged liquid, much like vibrating strings. Alfv\'en
initially believed that these waves played a role in the solar sun spot
cycle.\footnote{See \cite{Alfven1942} and
\cite[chap.~10]{Jackson1975}.} While this expectation has not been confirmed,
these kind of magnetohydrodynamic waves derived for charged liquids of
infinite conductivity nevertheless do play a role in plasma physics.

\section*{Microscopic theory of charge transport mechanism}
\addcontentsline{toc}{section}{Microscopic theory of charge transport mechanism}

In the preceding section, we have addressed superconductivity on a
phenomenological level exclusively as a special case of infinite
conductivity, i.e.\ as far as its implications in the framework of
Maxwell's equations go. But physicists at the time also entertained
speculations on a microscopic level, i.e.\ on the level of model
assumptions about superconductive mechanisms of electric charge transport.
In fact, investigations of the Hall effect were done to some extent because
the magnitude and sign of the Hall voltage carries information on the charge
carriers and especially on their sign. If specific experimental data had been
available on a Hall effect for superconductors, this would have had 
direct implications for speculations on the microscopic level.

The discussion of microscopic models of electric conductivity that we are going 
to discuss in the following were prompted by the phenomenon of superconductivity. 
To be sure, some of the models, or at least certain aspects of them,
were not necessarily new. But for the purposes of the present account, I will discuss 
the contemporary microscopic speculations only to the extent that is needed to 
establish the historical horizon for Einstein's own contributions in the period 
under consideration.
In particular, I will refrain from making any claims about the prehistory of
individual models of charge transport.\footnote{See 
also footnote \ref{note:disclaimer} above.}

\subsection*{Stark's model of thrust planes}
\addcontentsline{toc}{subsection}{Stark's model of thrust planes}

\begin{figure}[thb]
\begin{center}
\includegraphics[scale=1.00]{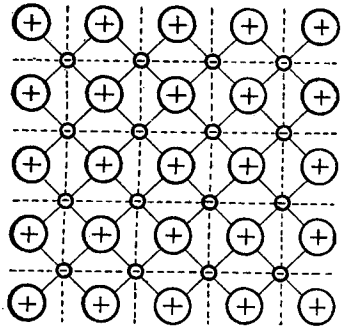}
\caption{A regular lattice of a monovalent metal with negative valence
electrons and positive spheres. Dashed lines denote intersections
between the paper plane and Stark's ``thrust planes'' along which aggregates of
valence electrons are supposed to move along in force-free motion
(from \cite[p.~192]{Stark1912}).}
\label{fig:Stark}
\end{center}
\end{figure}

Alternatives to Drude's electron theory of metals were advanced
in order to account for a number of unexplained experimental facts. 
Foremost among them was the
problem that the observed electric conductivities of metals would
imply an electron density that would also give an appreciable electronic
contribution to the specific heat. No such contribution, however, was
seen experimentally. With reference to this problem, Johannes Stark
published an alternative theory of electric conductivity in
1912.\footnote{\cite{Stark1912}.} It is of interest here because Stark
also alluded to the recently discovered superconductivity for its
justification. His theory is based on what he called a ``valence
hypothesis'' (``Valenzhypothese'') according to which ``point-like
separable, negative electrons are situated at the surface of the chemical
atoms vis-a-vis extended, inseparable positive spheres.''\footnote{%
ibid., p.~191.}
In metallic conductors, these valence electrons are located at some
distance away from the positive spheres. For monovalent metals the
negative electrons and the positive spheres may crystallize into a
regular lattice, as shown in Fig.~(\ref{fig:Stark}), where the solid
lines indicate lines of force between the electrons and the positive
spheres. A single electron cannot move about easily within such a
lattice aggregate, since local forces would immmediately pull it back
to its equilibrium position. But there are certain directions along
which electrons may move without doing work; these directions are
parallel to planes located symmetrically between the positive spheres,
such as those planes whose intersection with the paper plane is
indicated by the dashed lines in Fig.~(\ref{fig:Stark}). Stark calls
such planes ``thrust planes'' (``Schubfl\"achen''). Along those planes
a valence electron may be moved ``together with many other valence electrons
by arbitrarily small forces.''\footnote{%
ibid., p.~193.}
Electric resistivity for those collective motions then would arise
from thermal vibration of the positive spheres as well as from lattice
defects. It follows that perfect conductivity is possible in the limit
of zero absolute temperature. A metal that allows for such motion and 
electric conduction at absolute zero temperature is called by 
Stark a ``whole metal'' (``Ganzmetall''). Its resistance vanishes
at zero temperature and increases ``with increasing
number of valence fields that are
momentarily in a vibrating state and with increasing amplitude of
these vibrations.''\footnote{%
ibid., p.~194.}
In contrast to the standard Drude model, this theory accounts, at
least qualitatively, for the possibility of perfect conductivity 
but does not account for the sudden loss of
resistivity at a low but finite low transition temperature.
Accordingly, Stark refers to the recent Leiden findings on the
conductivity for mercury not as a sudden loss of resistivity but as a
limiting phenomenon for temperatures approaching absolute
zero.\footnote{``As they [i.e.\ the Leiden investigations] have shown,
the conductivity, especially that of mercury, does not increase towards a
maximum and then decreases again for decreasing temperature as a
result of the smaller number of free electrons, but it approaches even
infinity when one goes to absolute zero temperature.'' 
\cite[p.~191]{Stark1912}.}

\subsection*{Lindemann's model of electron space-lattices}
\addcontentsline{toc}{subsection}{Lindemann's model of electron space-lattices}

A similar model of collective motion of electrons that move about by
preserving some lattice structure was proposed a few years later, in 1915, by
F.A.~Lindemann. Lindemann also pointed to the difficulties of the free 
electron model, in particular to
the problem of the electronic specific heat. With reference to the
magnitude of the Coulomb forces that act between electrons at a typical
density in the metal, Lindemann argued that it is impossible to ignore
the interaction between the electrons:
\begin{quote}
The expression free electron, suggesting, and intending to suggest, an
electron normally not under the action of any force, like an atom in a
monatomic gas, might almost be called a contradiction in
terms.\footnote{\cite[p.~129]{Lindemann1915}.}
\end{quote}

Instead, he put forth the hypothesis that ``the electrons in a metal
may be looked upon as a perfect
solid.''\footnote{ibid.} Lindemann argued that
in addition to their mutual repulsion, electrons are attracted by
electrostatic forces to the positive ions up to a certain radius $r_0$ where
a repulsive force between the core of the ions and the electron sets
in. His model then amounted to the assumption that a ``metal crystal
would consist of two interleaved space-lattices, one consisting of
atoms or ions, one of
electrons.''\footnote{ibid., p.~130.} The details and
quantitative mathematical consequences of his model would be ``a
matter of great difficulty,'' Lindemann conceded. But he indicated that
he imagined
the whole electron space lattice to shift with respect to the atomic
lattice when an external field is applied, and that the 
electron space lattice may move continuously, with electrons at one
end leaving the lattice structure which would continuously be
filled up again at the other end, when a source of electrons is applied.
\begin{quote}
In other words, the electron
space-lattice or crystal may be said to melt at the one end and fresh
layers may be said to freeze on at the other end when a current
flows.\footnote{\cite[p.~130]{Lindemann1915}.}
\end{quote}
In order to account for superconductivity, Lindemann then argued that
as long as the radius of the repulsive ion core, $r_0$, is less than
half the distance between the centers of the atoms, ``the electron
space-lattice can move unimpeded through the atom space-lattice.''
Again, just as in Stark's theory, electric resistivity would set in
through thermal vibrations of the positive ions. But in contrast to
Stark's theory, this state of superconductivity would certainly be possible
at a low but finite temperature.

\subsection*{Thomson's model of electric dipole chains}
\addcontentsline{toc}{subsection}{Thomson's model of electric dipole chains}

Another early reaction to Onnes' discovery of superconductivity was
also published in 1915 by J.J.Thomson. According to Thomson,
Onnes' experiments, showing that the specific resistivity of some
metals drops to ``less than one hundred thousandth millionth part of
that at 0$^{\circ}$C,'' were of ``vital importance in the theory of
metallic conduction.''\footnote{\cite[p.~192]{Thomson1915}.} Thomson
was especially intrigued by the demonstration of the existence of
persistent currents. In addition to the apparently complete loss of
resistivity, Thomson emphasized the fact that the transition
takes place at a definite temperature and that the loss of
resistivity seemed to occur almost instantaneously. This fact seemed
to him to be another ``fatal objection'' to the model of free electrons.
With reference to his earlier work, Thomson now advanced a theory
of electric conduction based on the assumption that the main
mechanism of current transport is due to the existence of electric
dipoles or, in Thomson's words ``electrical doublets, i.e.\ pairs of
equal and opposite charges at a small distance
apart.''\footnote{ibid., p.~193.} The existence of these
doublets renders the substance polarizable, and Thomson proceeded to
develop a quantitative theory of the temperature dependence of the
electric polarization, in direct analogy to Langevin's calculation of
the magnetization on the basis of the kinetic theory of gases.

\begin{figure}[thb]
\begin{center}
\includegraphics[scale=0.50]{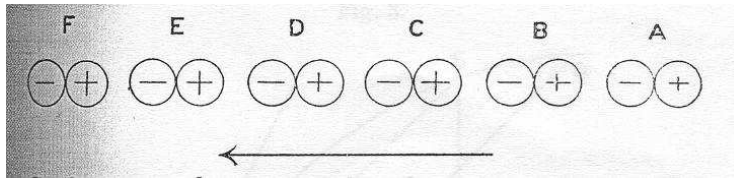}
\caption{A chain of aligned electric dipoles as imagined in Thomson's 
explanation of electric conductivity (from \cite[p.~195]{Thomson1915}).}
\label{fig:thomson1915}
\end{center}
\end{figure}

For any finite value of the electric polarization, Thomson
argued, we may assume that some of the doublets are
pointing into the same direction, while the rest of them are pointing
in random directions. Furthermore, Thomson suggested, ``picture the
substance as containing a number of chains of polarized atoms whose
doublets all point in the direction of the electric
force.''\footnote{ibid., p.~195.} To illustrate his
model, Thomson included a sketch of one such chain, as shown in
Fig.~\ref{fig:thomson1915}.
These considerations were valid generically for both insulators
and conductors.  The crucial point of Thomson's model was the
assumption that the motion of the conducting electrons is not affected
by the external electric force but rather by the local electric forces
of the atoms in the chain of doublets.
\begin{quote}
On this theory the peculiarity of metals is that electrons, not
necessarily nor probably those in the doublets, are very easily
attracted by these forces from the atoms when these are crowded together. 
Thus we may suppose that under these forces an electron is torn from $A$
and goes to $B$, another from $B$ going to $C$, and so on along the
line,---the electrons passing along the chain of atoms like a company
in single file passing over a series of
stepping-stones.\footnote{ibid., p.~195.}
\end{quote}
The conceptual distinction between the external electric force and the
local forces exerted by the doublets, which are the forces that are
actually acting on the conduction electrons, allows Thomson also to
account for the phenomenon of superconductivity:
\begin{quote}
[...] the part played by the electric force in metallic conduction is
to polarize the metal, i.e.\ to form chains: when once these are
formed the electricity is transmitted along them by the forces exerted
by the atoms on the electrons in their neighbours. Thus if the
polarization remains after the electric force is removed the current
will remain too, just as it did in Kamerlingh Onnes' experiment with
the lead ring.\footnote{ibid., p.~198.}
\end{quote}
A strong point of Thomson's theory thus is the analogy to the
ferromagnetic phase transition of paramagnets. Since it is the
polarization that accounts for the electric conductivity, the model
can explain, at least in principle, why the transition to the
superconducting state happens discontinuously.

\subsection*{Kamerlingh Onnes's model of superconducting filaments}
\addcontentsline{toc}{subsection}{Kamerlingh Onnes's model of superconducting
filaments}

Thomson's model was received favorably by Kamerlingh Onnes who slightly
modified it. At the 1921 Solvay Congress, Onnes gave a report on the state
of knowledge about superconductivity, in which he also included a
discussion of microscopic electronic
theories.\footnote{\cite[\S~5.]{Onnes1921b}. Einstein had been invited 
to attend the 1921 Solvay Congress and to talk about recent experiments 
on the gyromagnetic effect (see Lorentz to Einstein, 9 June 1920, 
\cite[Doc.~49]{CPAE10}) but decided instead to travel to the U.S.\ 
on a fundraising mission for the Hebrew University.}  As in Thomson's
analysis, Onnes emphasized two features of the phenomonenon which he
singled out as fundamental: the complete loss of resistivity and the
discontinuity of the transition. In view of the latter, Onnes asked
whether there would be any other quantity that would undergo a sudden
change at the superconductive transition and emphasized that there
appeared to be none. In particular, he emphasized that no
corresponding change of thermal conductivity was observed, and that
in the superconducting state there would be ``no longer any trace of
the law of Wiedemann and Franz.''\footnote{%
\cite[p.~45]{Onnes1921b}.}
An attractive feature of Thomson's model was that it could account for 
the discontinuity of the transition. But, wrote Onnes, with his hypothesis
of an alignment of the doublets and the molecular field thus created
Thomson went ``perhaps a bit too far'' in specializing his assumptions
than would be necessary to explain the discontinuous transition.\footnote{%
ibid., p.~46.}
Instead, Onnes wondered whether the conduction electrons had, in
general, ``two ways of moving about in the atomic
lattice.''
One way, above the transition
temperature, would be less ordered with frequent collisions with the
atoms, and another one more ordered would take place below the
transition temperature. Here the conduction electrons would ``slide,
by a sort of congelation, through the metallic lattice without hitting
the atoms.''
But Thomson's general idea was still good, i.e.\ the idea
``of a discontinuity determined by the temperature where some process
has the character of an alignment.''
Onnes discussed the difficulty of explaining the large mean free
paths needed to account for the loss of resistivity according to the
standard theory. He concluded that the notion of a mean free path has
to be abandoned and replaced by a related concept:
\begin{quote}
We assume that under certain circumstances filaments of great length
are being formed, along which an electron, that takes part in the
conduction, can glide on the surface of the atoms and pass from one
atom to the other without transmitting any energy to those degrees of
freedom that contribute to the statistical equilibrium of the thermal
motion.\footnote{%
ibid., p.~47.}
\end{quote}
Such motion would hence be called ``adiabatic''
(``adiabatique''). Those filaments need not, in contrast to Thomson's
model, be rectilinear but could be curved or twisted; they need not be
made up necessarily from the same sort of atoms and could have
ramifications everywhere, so that the electron might pass back and forth
along these filaments throughout its path, always following the
conditions of the superconductive state. 

The adiabatic motion would have to be complemented by some
non-adiabatic process at the ends of the filaments. As to the precise
nature of those non-adiabatic events, Onnes only ventured a few
conjectures in a footnote:
\begin{quote}
This could be the ejection of an electron from the atom, its passage
in the state of free motion, and its collision with another atom, or
else the immediate transport to an atom that comes into collision with
the end of the filament, or the rupture of the filament by thermal
agitation, if one lets oneself be guided by the old images, or else
some other process of transmitting the ordered energy of the electrons
to the thermal motion, if one strives to approach the theory of
quanta.\footnote{%
ibid., p.~48.}
\end{quote}

Kamerlingh Onnes also observed that the notion of a collision of an
electron would have to be generalized. The generalized notion would
mainly have to render understandable how an electron can pass on its
kinetic energy (``quantit\'e de mouvement'') to the thermal energy of
an atom. At this point, he added a footnote, alluding to a kind of
billiard ball mechanism of electronic collisions:
\begin{quote}
As soon as the superconducting state was discovered, one had observed
the analogy between the way in which the electricity is transported in
a superconductor and that in which, in a common experience that one
can do with a row of billard balls suspended one next to each other,
the momentum propagates from the first ball to the
last.\footnote{%
ibid.}
\end{quote}

Onnes remained vague at this point
as to the precise mechanism that would be responsible for
superconductivity. He referred in the end to the new theory of
quanta, and formulated as a task for research to find a model of
the atom that would allow a precise understanding of
``{\it this sort of electromagnetic crystallization,} that, below
a certain temperature, brings together all of a sudden the outer
electrons of a huge number of atoms into filaments of a macroscopic
order [...].''\footnote{%
ibid., p.~49.}

\subsection*{Haber's model of osculating quantum orbits}
\addcontentsline{toc}{subsection}{Haber's model of osculating quantum orbits}

The models of mechanisms of electric conduction discussed so far
were based exclusively on classical concepts and did not invoke any of
the new concepts associated with the emerging quantum theory. But by
1919, the success of the Bohr-Sommerfeld model of the atom suggested
that these concepts should also be exploited for an understanding of the
open problems in the theory of electric conductivity. This is
what Kamerlingh Onnes had asked for in his contribution to the 1921
Solvay Congress. Before proceeding to discuss Einstein's views on
these matters, we will discuss one such proposal to make use
of the new quantum theory of the atom for a new understanding of the
phenomenon of superconductivity made by Fritz Haber in an addendum 
to the second of two communications devoted to the theory of metallic
structure.\footnote{\cite{Haber1919a,Haber1919b}.}  

Haber attempted to come to a better
understanding of the structure and properties of metals by conceiving
them as being made up of regular lattice structures where the lattice
sites are occupied by positive ions and negative electrons and where
the lattice energy is computed taking into account both van-der-Waals
forces between the ions and the electrostatic forces between electrons
and ions. More specifically, Haber computed quantitative relations
between volume and compressibility on the one hand, and ionization
energy of the metallic atoms and the heat of sublimation on the other
hand and compared the theoretical values with observational data in
order to test his general hypothesis. Haber's second communication on the
subject was presented to the Prussian Academy for publication on 27
November 1919, and an addendum to the second communication was written
after Peter Debye had presented results about X-ray diffraction
studies of lithium to the German Chemical Society on 29 November. 
Debye had shown that only core electrons were detected at
the lattice sites of a body-centered cubic lattice, and that no valence
electrons were observed that would be located at fixed lattice sites
or on fixed orbits around lattice sites. As a consequence of these
findings, Haber modified his original proposal to the effect that only
positive ions make up the regular lattice structure, and that the outer
electrons orbit around the positive cores in the interstitial spaces. Haber
called the original model with both electrons and ions at the lattice
points the ``static picture of the metal'' (``das statische Bild des
Metalls.'')\footnote{\cite[p.~1002]{Haber1919b}.}. The case where the
lattice is only made up of positive core ions is called a ``dynamic 
lattice'' (``Bewegungsgitter''). Conceiving of metals as dynamic lattices
also solved, according to Haber, the difficulty posed by the
phenomenon of superconductivity:
\begin{quote}
If the electrons were sitting fixed in the lattice sites, it
could not be understood how the superconductivity at absolute zero
temperature came about without violation of Ohm's law. In this case a
minimal force would be needed to effect their translation from one
lattice point to the other.\footnote{%
ibid.}
\end{quote}
But if metals were ``dynamic lattices,'' one would also be able
to account for superconductivity. The idea was to invoke Bohr's concept
of stationary electron orbits around positive cores and to assume that electrons
may both move on these orbits and also, under certain conditions, pass
easily from one orbit to the other:
\begin{quote}
The point of view that naturally comes to mind is to conceive
of superconductivity as a state in which the valence electrons of the
metal move in orbits that have common tangents in points of equal
velocity.\footnote{%
ibid., p.~1003.}
\end{quote}
Since, according to Bohr's quantum hypothesis, the electrons move
around the atom cores on stationary orbits without radiating off
electromagnetic energy, they can thus move along from atom to atom
and give rise to a conduction current without electric resistivity.
Therefore, in a dynamic lattice, an electric current may flow if one applies
``an ever so weak field.''

Haber proceeded one step further to put his hypothesis to a
quantitative test. In a body-centered cubic lattice half the distance
between nearest neighbouring lattice sites is $r=\delta\sqrt{3}/4$
where $\delta$ is related to the molecular volume $V/N$ by
$\delta^3=2V/N$. Circular orbits around the lattice sites that would
have ``common tangents'' would hence have a radius of this value.
Haber now invoked Bohr's quantum condition for circular orbits, i.e.\
$\int pdq = mvr = nh/2\pi$ where $m$ is the electron's mass, $v$ its
speed, $h$ is Planck's constant, and $n$ the quantum number. From this
quantization condition, it follows that the electrons would have a
kinetic energy $mv^2/2$ that could be seen as the frequency $\nu_s$,
needed to kick out the electron in the photoelectric effect. Haber thus wrote
the quantum condition as\footnote{%
ibid., p.~1004.}
\begin{equation}
\frac{mv^2}{2}\cdot 2mr^2 = \frac{n^2h^2}{4\pi^2} = 
h\nu_s\frac{2^{2/3}V^{2/3}m}{N^{2/3}}\frac{3}{8},
\end{equation}
where the second equation now expresses a testable relation between
the empirically accessible quantities $\nu_s$, $V/N$, and $m$. Taking
$n=2$ for monovalent metals, Haber found ``a reasonable representation
of our experience for all monovalent metals, except for lithium and
sodium, where our idealized model obviously does not
suffice.''

\subsection*{Einstein's model of conduction chains}
\addcontentsline{toc}{subsection}{Einstein's model of conduction chains}

Einstein's reaction to this kind of speculation about charge transport
mechanisms on a microscopic level was characteristically twofold. He 
was a party to the debate and contributed an idea that was
actually put to an empirical test by Kamerlingh Onnes. He also
reflected on the theoretical situation from an epistemological point
of view. Let us discuss Einstein's own model first.

\begin{figure}[thb]
\begin{center}
\includegraphics[scale=0.30]{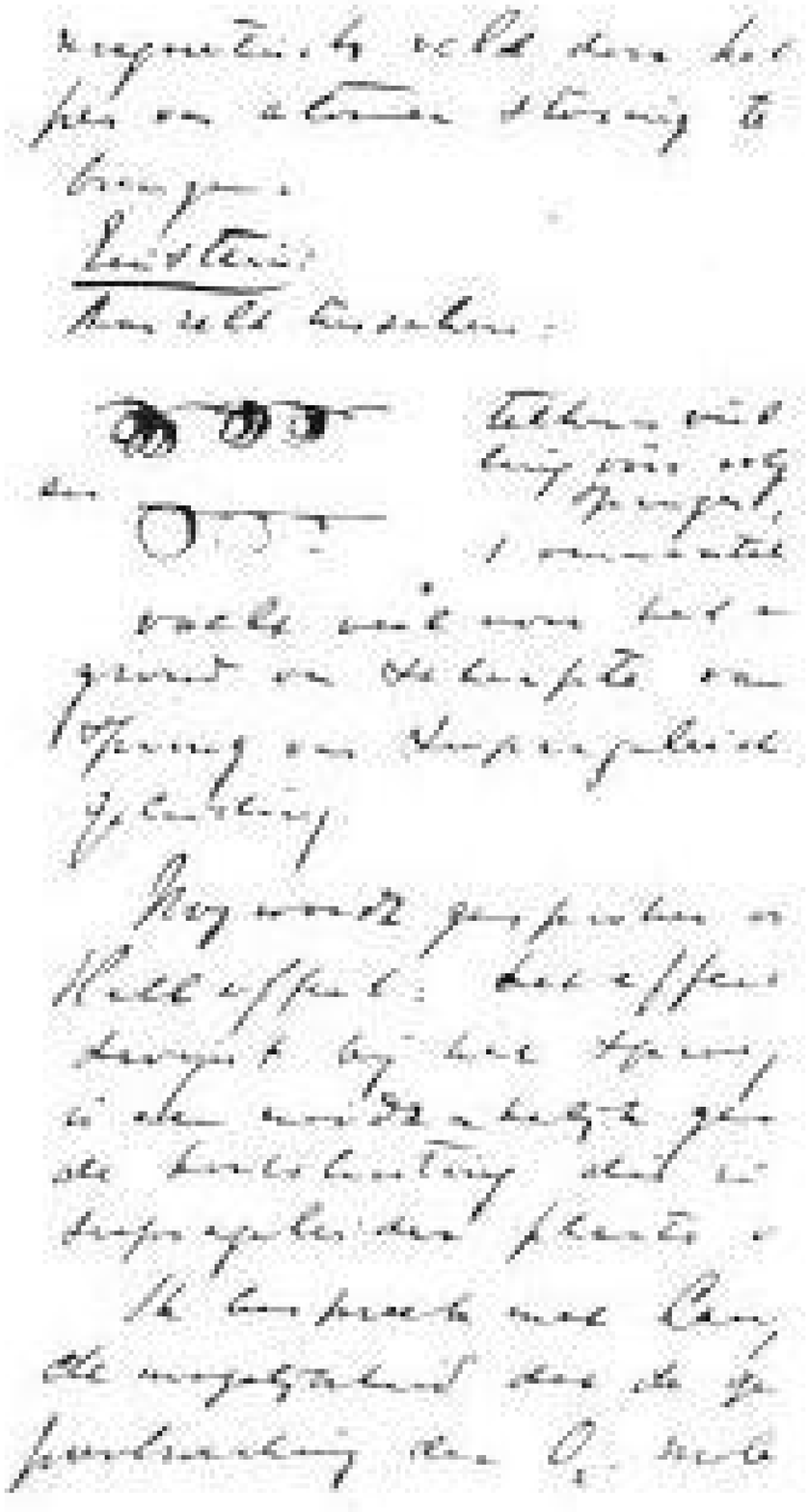}
\includegraphics[scale=0.30]{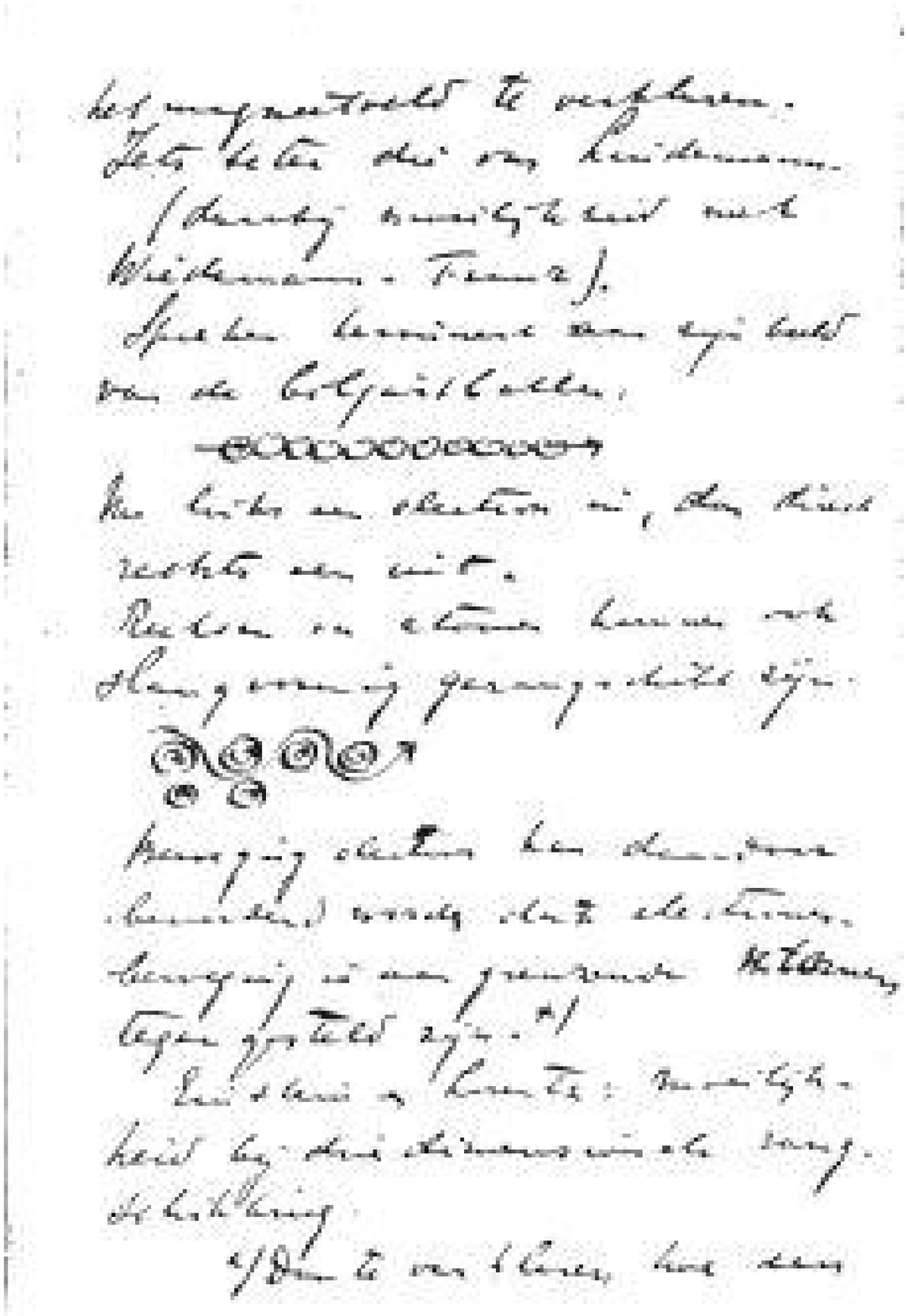}
\caption{Notes by Willem H.~Keesom about models of superconductivity
taken during discussions at the ``Magnet-Woche'' in Leiden in November 1919
(from \cite[p.~41]{MV2003}).}
\label{fig:keesom2}
\end{center}
\end{figure}

We have some indirect evidence that the phenomenon of
superconductivity was discussed not only phenomenologically but also
on the microscopic level during the Leiden ``Magnet-Woche'' in early
November 1919. The blackboard shown in Fig.~\ref{fig:blackboard}
appears to hold sketches of what may well be models of electron
trajectories. We also have some brief and sketchy notes by one of the
participants, Willem H.\ Keesom, that have been discussed and partly
reproduced in facsimile in \cite[pp.~41--42]{MV2003}, see
Fig.~\ref{fig:keesom2}.
The sketches in those figures suggest that the participants discussed
models of superconductivity similar to Fritz Haber's theory. Indeed,
the notes by Keesom indicate that Einstein seems to have been debating
whether an electron would revolve many times around an atom
before making the next jump, or whether it would revolve only once. It
seems that he was inclined toward the former case
in view of the sharpness of the transition between normal
conductivity and superconductivity. 

\begin{figure}[thb]
\begin{center}
\includegraphics[scale=0.50]{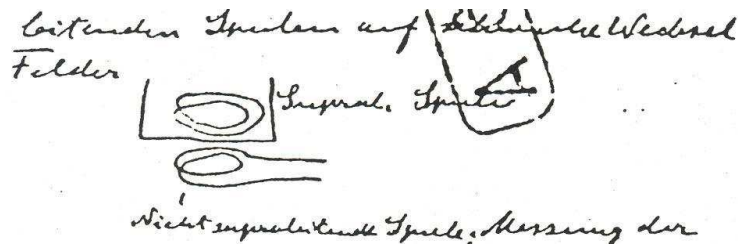}
\caption{Sketch of proposed experiment in Einstein's letter to Paul Ehrenfest, 
2 November 1921, AEA~9-566}
\label{fig:AEcoils}
\end{center}
\end{figure}

A year later, Einstein entertained some
concrete ideas along these same lines. In a letter to Paul Ehrenfest,
dated 2 November 1921, Einstein again picked up the topic of a
microscopic theory of superconductivity.
\begin{quote}
Do you remember our discussions about the superconductor? I am getting
back to this again. If there are no free electrons in the metals, then
an electric current means that there are electrons whose well-ordered
trajectory goes from atom to atom, and in the case of
superconductivity it does so in a stationary way. But it cannot be single
electrons because of the electric incompressibility. Hence it must be
electron chains that are formed by atom-electrons marching in single
file as it were. These chains are permanent and undisrupted in the
state of superconductivity.\footnote{%
Einstein to Paul Ehrenfest, 2 September
1921, AEA~9-566.}
\end{quote}
So far, Einstein's idea is strongly reminiscent of J.J.~Thomson's
model, although he did not invoke the idea of electric dipoles but
instead referred to ``atom-electrons'' (``Atom-Elektronen''). 
Einstein continued to draw some immediate
consequences from his hypothesis. He assumed that an
electric current is only possible through a chain that extends over
the entire substance between two points. Each chain of conduction 
electrons extending between two points contributes one unit of current.
\begin{quote}
The current is proportional to the number of such chains, hence it can
take on only discrete values.\footnote{%
ibid.}
\end{quote}
One such unit would be given
by the charge of an electron times the velocity with which it is
moving in those chains:
\begin{quote}
The discrete quantity of current is of the order $\nu e$ (opt[ical]
frequency $\cdot$ charge of the electron).\footnote{%
ibid.}
\end{quote}
This suggests that Einstein was thinking more along the
lines of Haber's model. The optical frequency refers to the circular
frequency of an electron travelling around the atom on a quantum
orbit.\footnote{In Bohr's atomic model the circular frequency $\omega$
is of order $\omega=h/(2\pi m_er^2)=2\pi\nu$ where $h=6.6\cdot 10^{-34}$Js,
$m=9.1\cdot 10^{-31}$kg, and $r\geq 0.5\cdot 10^{-10}$m, hence
$\nu\lessapprox 7.3\cdot 10^{15}$/s, for ground-state hydrogen and 
smaller for outer orbits of larger atoms. The human eye is sensitive 
to electromagnetic radiation in the frequency range 
$\nu \approx 0.75\dots 0.43\cdot 10^{15}$/s.} Einstein now invoked a genuinely
non-classical feature of the new quantum model, i.e.\ the assumption 
that electrons move on quantum orbits with discretely defined momenta.
\begin{quote}
If this is correct, {\it then a superconducting coil would not respond
to arbitrarily small electromotive forces}, hence would not screen
magnetic fields that are brought about sufficiently slowly (and that
are weak enough so as not to destroy the superconductivity). The
expression ``superconductivity'' would then be
misleading.\footnote{%
ibid.}
\end{quote}
The point is that since the superconducting current can only flow
along the chains, and since the electrons travel on the quantized orbits,
their velocity is fixed by the quantum conditions of Bohr's atomic
model. Consequently, there should be a finite minimal electric current that
must be excited. Einstein suggested that this consequence
should be put to experimental test in Leiden:
\begin{quote}
Such an experiment should be performed by you. 
[...] The superconducting coil could not
carry currents below $10^{-4}$ up to $10^{-5}$ Amp\`ere. Stronger
magnetic fields destroy the chains.\footnote{%
ibid.}
\end{quote}
More concretely, Einstein suggested measuring the self-induction of a
non-superconducting coil that is placed next to a superconducting
one, see Fig.~(\ref{fig:AEcoils}).
If the superconducting coil could take on only discrete and finite
values of current, this feature should show up in the apparent
self-induction of the non-superconducting coil.  The minimal value of
a superconducting current quoted by Einstein follows readily from his
assumption that the circular frequency of the orbiting electrons is in
the optical range. Indeed, the product of $\nu\cdot e$ evaluates to
$\approx 1.5\cdot 10^{-5}$A if we take $\omega\approx 10^{16}/s$.

The occasion for Einstein's returning again to the problem of
superconductivity may well have been an invitation to contribute to a
{\it Gedenkboek} to be published on the occasion of the fortieth
anniversary of Onnes's appointment as professor in 
Leiden.\footnote{\cite{Onnes1922}.} A direct
response by Ehrenfest to Einstein's letter is missing or not extant but he may well
have alerted Einstein to the fact that his model was reminiscent of
some ideas existing in the literature. In a letter to Ehrenfest,
written about two months later, Einstein referred to what is probably
his contribution to the Onnes {\it Gedenkboek}:
\begin{quote}
[I am] Citing Haber in my article on superconductivity. He had developed a
similar conception a few years ago in an Academy paper, albeit without
``snakes.''\footnote{%
Einstein to Paul Ehrenfest, 11 January 1922, AEA~10-004.}
\end{quote}
Einstein's published contribution to the {\it
Gedenkboek}\footnote{\cite{Einstein1922}.} contains an
explicit reference to Haber's 1919 paper discussed in the
previous section.\footnote{The reference was made in the postscript
and refers to the first page of \cite{Haber1919a} rather than more
specifically to the addendum to \cite{Haber1919b}.} 
After arguing that there cannot be any free
electrons in a metal, he continued with his hypothesis about metallic conduction.
\begin{quote}
Then metallic conductivity is caused by atoms exchanging their
peripheral electrons. If an atom received an electron from a
neighboring atom without giving an electron to another neighboring
atom at the same time it would suffer from gigantic energetic changes
which cannot occur in conserved superconducting currents without
expenses in energy. It seems unavoidable that superconducting currents
are carried by closed chains of molecules (conduction chains) whose
electrons endure ongoing cyclic changes.\footnote{%
\cite[pp.~433--434]{Einstein1922}.}
\end{quote}
In contrast to Haber's discussion, Einstein here emphasized that the
electrons would have to move collectively in ``conduction chains,'' 
much like in Thomson's
model. This in any case seems to be the sense of his remark that
Haber did not have the idea of ``snakes.'' In the published version,
Einstein hardly was any more specific about his model of electric
conduction.
But he did repeat his suggestion to test the implication
of a finite current threshold for superconductors.
\begin{quote}
[...] there is the possibility that conduction chains cannot carry
arbitrarily small currents but only currents with a certain finite
value. This would also be accessible to experimental
verification.\footnote{%
ibid., p.~434.}
\end{quote}
This experiment seems not to have been done in Leiden. But
another consequence of his model that he proposed for experimental
investigation was tested explicitly in an experiment done by
Kamerlingh Onnes. Einstein's idea of ``conduction chains'' along
atomic quantum orbits was restrictive not only because
it allowed only for quantized units of current. It was also
restrictive in the sense that it did not allow for chains to be made up
of different atoms, since the orbital velocities around different atoms
would differ, and hence would not allow for smooth transitions of the
conduction electrons from orbit to orbit.
\begin{quote}
It may be seen unlikely that different atoms form conduction chains
with each other. Perhaps the transition from one superconducting metal
to a different one is never superconducting.\footnote{%
ibid.
}
\end{quote}
Einstein further argued for this model of conduction chains by
pointing out that it was quite natural that these chains would be
destroyed by large magnetic fields, as well as by thermal motion ``if
it is strong enough and if the $h\nu$ energy quanta that are being
created are big enough.''
Hence, it would also be understandable why superconductors turn into
normal conductors by raising the temperature, and one could understand
``maybe even the sharp temperature limit of
superconductors.''
Indeed, Einstein conjectured that normal electric conductivity may 
perhaps be nothing else but
superconductivity that is constantly being destroyed by thermal
motion. This conjecture, he concluded, would be suggested by the
``consideration that the frequency of the transition of the electrons
to the neighboring atom should be closely related to the circulation
frequency of electrons in the isolated atom.''\footnote{%
ibid., p.~435.} 
The very last sentence of his paper then repeats the hypothesis that 
superconductors must necessarily be homogeneous:
\begin{quote}
If this idea of elementary currents caused by quanta proves correct it
will be evident that such chains can never contain different
atoms.\footnote{%
ibid.}
\end{quote}

We have reason to believe that Einstein was eager to see whether these
consequences would actually be observed. On 21 January 1922, he wrote
to Ehrenfest:
\begin{quote}
Nurture Onnes about those
superconductivity-experiments.\footnote{``Sch\"ure Onnes wegen der
Supraleitungs-Versuche.'' Einstein to Paul Ehrenfest, 21
January 1922, AEA~10-011.}
\end{quote}
Indeed, a few weeks later, Ehrenfest reported back to Einstein that
Onnes had investigated the issue of whether the interface between
different superconducting materials would still be superconductive, and
that he had found that no resistance was observed for a contact
between tin and lead.\footnote{Paul Ehrenfest to Einstein, 11 March
1922, AEA~10-025.} Ehrenfest added that Onnes would write to Einstein
himself about these findings, but that letter seems to have been lost. In any
case, Einstein added a postscript to his {\it Gedenkboek}
contribution. Referring to his final remark on the impossibility of 
having conduction chains contain different atoms, he added:
\begin{quote}
The last speculation (which by the way is not new) is
contradicted by an important experiment which was conducted by
Kamerlingh Onnes in the last couple of months. He showed that at the
interface between two superconductors (lead and tin) no measurable Ohm
resistance appears.\footnote{%
\cite[p.~435]{Einstein1922}.
}
\end{quote}
It appears that the results of these experiments were never
published. But two years later, the very same experiment was repeated
with greater accuracy by Kamerlingh Onnes together with his student 
Willem Tuyn. The better accuracy was made possible by two modifications
of the experimental setup. For one, Onnes and his
collaborators had succeeded in isolating the liquid helium in a cryostat
that could be physically removed from the liquifier and transported to
a different location. They also employed a new method of determining
residual resistances by looking at persistent currents in rings, rather
than measuring the resistivity of filaments by directly observing the 
potential difference for strong currents. Details of these
experiments were presented by
Kamerlingh Onnes to the fourth Solvay Congress in April 1924, and
to the Fourth International Congress of Refrigeration, held in London
in June 1924.\footnote{\cite{Onnes1924}.} In the published report,
Onnes gave an overview of recent experiments and investigations into
superconductivity and discussed in section \S~5. ``diverse issues''
(``questions diverses'').

One of these was Einstein's hypothesis. Onnes began by mentioning that
he had shown ``with the method of filament''\footnote{%
ibid., p.~15.} that the resistance of
the ``soldered interface'' (``soudoure'') between lead and tin was
below what could be determined with the given limits of experimental
accuracy. Referring to Einstein's contribution to the {\it
Gedenkboek}, he remarked that Einstein had given up on ``his idea
that superconducting circuits cannot be constituted by different
atoms.''
He continued
\begin{quote}
Now that we have at our disposal a method for measuring these small
resistances with a much larger precision, it was of highest interest
to repeat these experiments.\footnote{%
ibid., p.~16.}
\end{quote}

The microresidual resistance was measured by the
lifetime of persistent currents in superconducting
rings. Specifically, Onnes used a ring of lead suspended on a
torsion rod within another slightly larger lead ring. The whole
setup was isolated against mechanical vibrations by mounting on a
shock absorber and immersed into liquid helium. Currents were induced
by an external magnetic field perpendicular to the plane of the rings,
and the inner ring was rotated out of its equilibrium position by an
amount of $30^{o}$. Afterwards, the motion of the inner ring was
monitored by light reflected from a mirror fixed to the torsion
rod. Since the currents were persistent, no rotational motion was
observed, and the setup gave an upper limit to the resistivity that was
determined by the time that the experiment could be run before the
liquid helium evaporated, a time that according to Onnes's report
took some six hours or so.

\begin{figure}[thb]
\begin{center}
\includegraphics[scale=0.30]{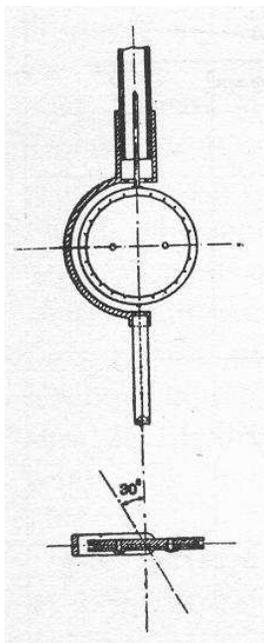}
\caption{Kamerlingh Onnes' experimental setup to test Einstein's hypothesis 
that junctions between two different superconducting metals are not 
superconductive (from \cite[p.~16]{Onnes1924}).}
\label{fig:onnes1924-fig7}
\end{center}
\end{figure}

In order to test Einstein's hypothesis,
Tuyn and Onnes now used a ring consisting of 24 alternating sectors of 
lead and tin, see Fig.~(\ref{fig:onnes1924-fig7}).
More precisely, the sectors consisted of bands of tin or lead wrapped
around a ring of ivory. Care was taken to keep the middle of each
sector cooled when soldering the sectors together in order to avoid
any diffusion of lead into tin. A current was induced in the ring by a
magnetic field, and the ring was displaced by an angle of
$30^{o}$. The expectation according to Einstein's hypothesis was that
it would take a certain amount of time for the current to die
down. However, the results did not accord with expectations.
\begin{quote}
But the experiment has shown that the currents continue to flow in the
ring and when the experiment was repeated when the ring was cut it
showed the same magnetic moment.\footnote{%
\cite[p.~16]{Onnes1924}.}
\end{quote}

This result was puzzling.\footnote{Fig.~\ref{fig:onnes1924-fig7} seems to show a
slightly different setup than was described earlier for the persistent
current measurements. Here only one half of the outer ring is shown.
This different is not commented on in Onnes's paper.}
Onnes presented his experiment as work in progress. Otherwise, he
argued, Einstein's hypothesis would have been proven:

\begin{quote}
Otherwise, one would already be driven to the conclusion that the 24
points of contact between the sectors have a resistance that is too
big to be measured by this method, since the current induced in the
complete circuit of the ring decays too rapidly alongside the
persistent currents induced in the individual
sectors.\footnote{%
\cite[p.~16]{Onnes1924}.
}
\end{quote}
Unfortunately, Onnes's description is not sufficiently detailed to allow 
an interpretation of the outcome of his experiments from our modern
understanding. If the interfaces between the sectors were clean, the
sectored ring should have shown a persistent current. If isolating
material had been added between the sectors, these would, in
principle, become tunneling barriers for the superconducting wave
function and the setup might perhaps have exhibited Josephson current 
effects. As described by Onnes, the experiments remain inconclusive.%
\footnote{In fact, in 1926 Einstein suggested to investigate this 
question once more in the low temperature laboratory of the 
{\it Physikalisch-Technische Reichsanstalt} (PTR) in Berlin.
In the discussion in its {\it Kuratorium}
following the presentation of the annual report of the PTR
for the year 1925 (when experiments at liquid helium temperatures had finally 
become possible), Einstein remarked that ``the question is of particular interest
whether the interface between two superconductors would be superconductive as well.''
See ``Bericht \"uber die T\"atigkeit der Physikalisch-Technischen Reichsanstalt
im Jahre 1925,'' copy deposited in the Library of the PTR, and minutes of the 
meeting of the {\it Kuratorium} of the PTR of 11 March 1926, Library of the PTR, 
sign. 240.2-241 (AEA~81-887), see also \cite[p.~95]{Hoffmann1980}. Einstein's
suggestion apparently was followed up on, but met with difficulties. In the report 
for the following year (1926), the authors wrote: ``The fact that alloys
become superconductive, makes it more difficult to decide experimentally the question,
posed by Einstein, whether a resistance appears at the interface of two superconductors
due to a breaking up of the superconductive conduction chains.'' \cite[p.~234]{Report1926}.} 

In the last section, Onnes discussed ``the structure
of superconductors,'' and again referred to Einstein:
\begin{quote}
I have accepted Einstein's idea that the electrons that take part in
the conductivity of a solid metal have velocities of the same order as
the valence electrons in the free atoms [...]\footnote{%
\cite[p.~26]{Onnes1924}.}
\end{quote}
As an immediate consequence of this assumption, the
melting transition of a metal should have little influence on the
conductivity, as Einstein had conjectured.\footnote{%
\cite[p.~433]{Einstein1922}.} But on a more
general level, it meant that the atomic model of the emerging quantum
theory had to be taken seriously for a theory of
superconductivity. For this reason, Onnes turned to Hendrik Anton
Kramers in Copenhagen, who 
provided him with a graphic visualization of the electronic structure of 
some of the metals under consideration, e.g.\
of Indium, as shown in Fig.~{\ref{fig:onnes1924-fig11}.
\begin{figure}[thb]
\begin{center}
\includegraphics[scale=0.40]{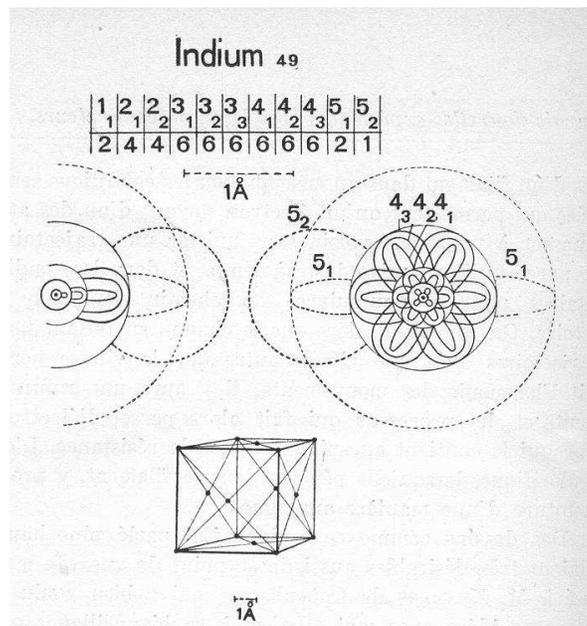}
\caption{Graph of the electronic and lattice structure of Indium, 
according to the Bohr-Sommerfled quantum theory (from \cite[p.~28]{Onnes1924}).}
\label{fig:onnes1924-fig11}
\end{center}
\end{figure}
Although Onnes went into some detail regarding the atomic
structure of metals and the consequences for a theoretical
understanding of superconductivity, his results remained
inconclusive as far as any quantitative results 
are concerned. Anything else would have been rather surprising from our
modern understanding of the phenomenon. Nevertheless, it is remarkable
not only that the phenomenon of superconductivity was perceived as a
genuine quantum phenomenon, but also that Einstein was among those,
who like Haber and Onnes,
clearly advocated making use of the Bohr-Sommerfeld theory for an
understanding of superconductivity.

In this context, another entry of around June 1922 in Ehrenfest's diaries is 
of some interest, see Fig.~\ref{fig:enb_bohr},
\begin{figure}[thb]
\begin{center}
\includegraphics[scale=0.40]{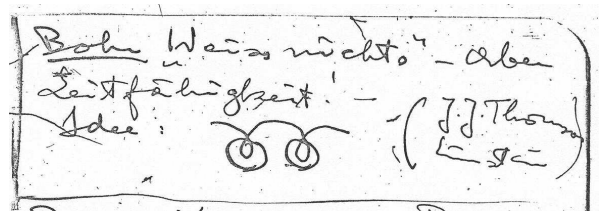}
\caption{A brief entry on an ``idea'' for conductivity with the names of Bohr, Einstein, and
Thomson in Ehrenfest's diaries. (ENB~4-19)}
\label{fig:enb_bohr}
\end{center}
\end{figure}
which suggests that Ehrenfest had
talked to Bohr himself about the issue. The entry says: ``Bohr:
``Don't know'' - but conductivity!  --- idea:'' and is accompanied by
a small sketch strongly reminiscent of Einstein's
conduction chains. Ehrenfest added in brackets the names of
J.J.~Thomson and Einstein. Whatever the context of this entry, it
supports the general conclusion that superconductivity was not only
investigated experimentally in Leiden, but also interpreted as part of
a larger attempt to come to an understanding of the new quantum
theory. 

It is in this sense that Onnes concluded his 1924 report 
by writing:
\begin{quote}
For the moment, in view of the state of the theory of quanta, it seems
that it would be utterly premature if one wanted to form more detailed
images, as I had in mind, of the motion of conduction electrons.\\
But one sees the dawning of the light that the application of this theory 
will bring.\footnote{%
\cite[p.~34]{Onnes1924}.}
\end{quote}

\section*{Einstein's epistemological reflections}
\addcontentsline{toc}{section}{Einstein's epistemological reflections}

So far, we have only discussed Einstein's comments on and considerations
about superconductivity as an attempt of a
contemporary physicist to come to a theoretical understanding of the
new phenomenon. But his published paper on the subject also carries a
distinctly and characteristically different overtone. In addition to
presenting and defending his own model speculation on a conduction
mechanism, it also offered quite explicit epistemological reflections on
the status of physical theory. Indeed, it begins like this:
\begin{quote}
The theoretically working scientist is not to be envied, because
nature, or more precisely: the experiment, is a relentless and not
very friendly judge [Richterin] of his work. In the best cases, she only says
``maybe'' to a theory, but never ``yes,'' and in most cases she says
``no.'' If an experiment agrees with a theory it means ``perhaps'' for
the latter. If it does not agree, it means ``no.'' Almost any theory
will experience a ``no'' at one point in time - most theories very
soon after they have been developed.\footnote{%
\cite[p.~429]{Einstein1922}.}
\end{quote}

Einstein had expressed similar
falsificationist views in a little piece on ``Induction and Deduction
in physics''\footnote{\cite{Einstein1919}.} 
published in the daily {\it Berliner Tageblatt} just some two years 
earlier in late 1919,
after the observational confirmation of gravitational light bending. 
There he argued that 
progress in physical theory usually does not occur by induction from 
empirical data but rather along some kind of hypothetico-deductive 
reasoning. The researcher, he wrote,
\begin{quote}
does not find his system of ideas in a methodical, inductive way;
rather, he adapts to the facts by intuitive selection among the
conceivable theories that are based upon axioms.\footnote{%
\cite[p.~219]{CPAE7}.}
\end{quote}
The experiment then appears, indeed, as a judge, and Einstein had
continued in a very similar way as in 1922 by expressing his
falsificationist leanings:
\begin{quote}
Thus, a theory can very well be found to be incorrect if there is a
logical error in its deduction, or found to be off the mark if a fact
is not in consonance with one of its conclusions. But {\it the truth}
of a theory can never be proven. For one never knows if future
experience will contradict its conclusion; [...]\footnote{%
ibid.}
\end{quote}
However, there is a subtle difference between Einstein's 1919 reflections
and those of 1922. In 1919, he was under the spell of the spectacular
confirmation of his most significant theoretical achievement, the
observation of the gravitational light bending, predicted by general
relativity.\footnote{For an account
of the expedition, its results, and Einstein's reaction to it, see
\cite[pp.~xxxi--xxxvii]{CPAE9}.} In 1922, Einstein reflected, as we
will see, on the failure of Drude's electron theory of metals in
light, or should one say, in darkness of the fact that no convincing
alternative was available to account for superconductivity. Hence, in
1919 he wrote that one never knows whether a theory will be proven
wrong by contradicting experience, while in 1922 he asserted that
``almost any theory will be proven wrong at some time.''

The justification for his epistemological pessimism was given in
Einstein's reflections on the present state of the theoretical
understanding of metallic conductivity. His point of departure is
Drude's electron theory of metals. He quoted Drude's formula for the
specific resistance $\omega$ of metals, i.e. the inverse of
eq.~(\ref{eq:sigma})\footnote{With the same problematic factor of $2$,
that was discussed above (see the discussion following eq.~(\ref{eq:Lorenz})).} 
\begin{equation}
\omega = \frac{2m}{\epsilon^2} \frac{u}{nl},
\end{equation}
where $m$ is the electron's mass, $\epsilon$ its charge, $u$ its mean
velocity, $n$ the electron density, and $l$ the mean free path, and
proceeeded to discuss the evidence against Drude's theory.

The difficulties arise from the implicit consequences of the
temperature dependencies of the mean velocity $u$, the electron density
$n$, and the mean free path $l$. The temperature dependence of 
the mean velocity is determined by the equipartition theorem
\begin{equation}
mu^2 = 3kT
\end{equation}
where $k$ is Boltzmann's constant, and $T$ the absolute
temperature.\footnote{In \cite{Einstein1922}, the factor of $3$ was
written erroneously on the left hand side of the equation
(cp.\ eq.~(\ref{eq:G}) above).}  Einstein
now argued that one might expect the electron density $n$ to increase
with temperature on the assumption that free conduction electrons are
created by thermally enhanced dissociation. But the resistance of 
metals typically increases
with temperature, rather than decreases. Hence, one might be tempted to
assume that $n$ is roughly temperature independent, and that some
temperature dependence of the mean free path arises from the thermal
lattice vibrations. But the first hypothesis would be problematic, and
the second one might be hard to justify quantitatively. Moreover, if
the mean free path is determined by the thermal energy of the metal,
one should expect that the resistance of non-superconducting metals
tends to zero for decreasing temperature, while in fact it remains
constant. The residual resistance might be explained by impurities, but
the effect of impurities on the mean free path would be to add a
constant to $1/l$. This, however, would change the resistance by an
amount proportional to $u$. But since the effect of impurities is to
change the resistance by a constant amount, one would have to assume
that $u$ does not depend on temperature. But, concluded Einstein,
\begin{quote}
under no circumstances can $u$ be assumed to be
temperature-independent, because otherwise the only success of the
theory, i.e.\ the explanation of the Wiedemann-Franz law, would have
to be sacrificed.\footnote{%
\cite[p.~432]{Einstein1922}.}
\end{quote}
The bottom line of Einstein's reflections on the implications of
Drude's result is that the thermal electron theory already fails to
account for the empirical facts of normal electric conductivity.
\begin{quote}
The breakdown of the theory became entirely obvious after the 
discovery of the superconductivity of metals.\footnote{%
ibid.}
\end{quote}
But since it was conceivable that the Wiedemann-Franz law might be
explained also by some other theory, Einstein retracted his
pessimistic epistemologic turn, if only vaguely.
\begin{quote}
No matter how the theory of electron conductivity may develop in the
future, one main aspect of this theory may remain valid for good,
namely the hypothesis that electric conductivity is based on the
motion of electrons.\footnote{%
ibid., p.~430.}
\end{quote}

Einstein's discussion of the epistemological status of a physical
theory against its empirical content may have been motivated only by
the wish to justify the putting into print of theoretical
speculations that until now he had aired only in personal
discussions, correspondence, and unpublished manuscripts. In any case,
he went on to present and justify his model of conduction chains,
based on conduction electrons that move on quantized atomic orbits. He
did emphasize that he considered these ideas little more than speculations:
\begin{quote}
Given our ignorance of the quantum mechanics of composite systems 
we are far away from being able to convert this vague idea into a
theory.\footnote{%
ibid., p.~434.}
\end{quote}
It is interesting that Einstein referred to the emerging
quantum mechanics of composite systems (``Quanten-Mechanik zusammengesetzter 
Systeme'') in this caveat. As we have seen,
his approach to a microscopic theory of superconductivity was
characteristically bold in putting these new concepts to use.
Incidentally, as conjectured by H.~Kragh, this may well be the first
time ever that the term ``quantum mechanics''
appeared in print.\footnote{\cite[p.~86]{Kragh1999}.} In any case,
Einstein's 1922 contribution encouraged the exploration
of new paths in the theoretical understanding of superconductivity.
\begin{quote}
This phantasizing can only be excused by the momentary quandary of the
theory. It is obvious that new ways of doing justice to the facts of
superconductivity have to be found.\footnote{%
\cite[pp.~434--435]{Einstein1922}.}
\end{quote}

\section*{Concluding remarks}
\addcontentsline{toc}{section}{Concluding remarks}

In this paper, I have argued that Einstein's appointment as a special
visiting professor at the University of Leiden in 1920 was motivated
to a considerable extent, if not primarily, by the fact that his 
Dutch colleagues perceived him to be a
leading theoretician of condensed matter physics, and especially of
low temperature physics. It was expected that he would contribute to
the theoretical understanding of new phenomena observed in the low
temperature regime, and that he would provide theoretical guidance to
experimental investigations undertaken in Leiden. It has also become
clear that Einstein himself tried to live up to these expectations, at
least during the period of time that we have been considering, 1919--1922. 

In his theoretical analyses of superconductivity, Einstein proposed at
least three experiments to be done in Leiden. His exploration of the
implications of Maxwell's equations for the case of perfect
conductivity led him to suggest a Hall experiment on a superconducting
slab of varying thickness. His proposal of conduction chains as a
microscopic mechanism of superconducting charge transport implied that
superconductive currents were quantized in magnitude and, in
particular, would show a minimal threshold value. He suggested
that this implication be tested by measuring the effective
self-induction of a coil of non-superconductive metal that was in
inductive contact with a superconducting coil. Another consequence of
his model was the implication that the interface between two different
superconductors would not be superconducting. This latter hypothesis
was explicitly tested by Onnes, with a negative result. The experiment
was repeated two years later with an experimental setup that allowed
for better accuracy but then produced results that were inconclusive.

It seems also fair to say that in the context of contemporary
theorizing about superconductivity, Einstein's considerations and
ideas were rather sophisticated and advanced. His exploration of the
implications of Maxwell's equations for perfect conductivity went well
beyond Lippmann's investigations and also proved to be more successful
and insightful than explorations along the same lines done by
Ehrenfest. Similarly, his microscopic model of conduction chains was
distinguished from alternative theories in that it went farthest in
the application of concepts of the emerging quantum theory for an
understanding of superconductivity.

One may regret that Einstein's thoughts about superconductivity
produced only one publication. But, from today's point of view, it is
also clear that, in spite of Einstein's insights and creativity, none of
his ideas would have brought about a better understanding of
superconductivity or of quantum physics for that matter.  The story of
Einstein's concerns with the phenomenon of superconductivity is hence
neither one of failure, nor is it one of success.  It is rather a
reflection of a peculiar situation of the state of theoretical physics
at the time that was characterized by an emerging division between
theory and experimental practice, and the fact that
the emerging quantum theory had not yet reached a stable and
convincing status. Einstein's falsificationist reflections on physical
theory vis-\`a-vis experimental observation seem to reflect the
division of labor that was embodied in his own status as a
theoretician for the Leiden cryogenic laboratory. His epistemological
pessimism was justified at the time in view of the weakness of quantum
theory. It is all the more suprising that he advocated so expressly an
application of quantum concepts for the theoretical understanding of
superconductivity.

\section*{Acknowledgments}

This paper owes its existence to Issachar Unna's discovery of the 
excerpts of Einstein's letter in Ehrenfest's diaries and to his and J\'oszef 
Illy's identification of the Burndy library manuscript page 
as another document relating to the same problem. It was the discovery
of these documents that gave me the idea to write this paper.
I am grateful to Jed Buchwald, Dieter Hoffmann, A.J.~Kox, Issachar Unna, 
Jeroen van Dongen, and especially to Diana Buchwald for a critical 
reading of earlier versions of this article and for valuable comments. I also wish 
to thank Giuseppe Castagnetti for providing information about the {\it
Physikalisch-Technische Reichsanstalt}, Georges Waysand for sending me 
copies of Keesom's notes, and Rosy Meiron for her help with some subtleties 
of the French language. Unpublished correspondence by Einstein is quoted 
with kind permission by the Albert Einstein Archives, The Hebrew University
of Jerusalem.

\end{document}